\documentclass[journal]{vgtc}                     

\onlineid{0}

\vgtccategory{Research}

\newcommand{\papertitle}{Touching or Chatting: The Utility of LLMs and Tactile Charts for Learning about Complex Chart Types by BLV Individuals}
\title{\papertitle}

\author{%
  \authororcid{Tingying He}{0000-0002-9670-5587},
  \authororcid{Maggie McCracken}{0009-0006-5280-0546.},
  \authororcid{Daniel Hajas}{0000-0002-2811-1197},
  \authororcid{Sarah Creem-Regehr}{0000-0001-7740-1118},
  \authororcid{Alexander Lex}{0000-0001-6930-5468}
}

\authorfooter{
    \item T. He (\raisebox{-.5pt}{\includegraphics[height=6pt]{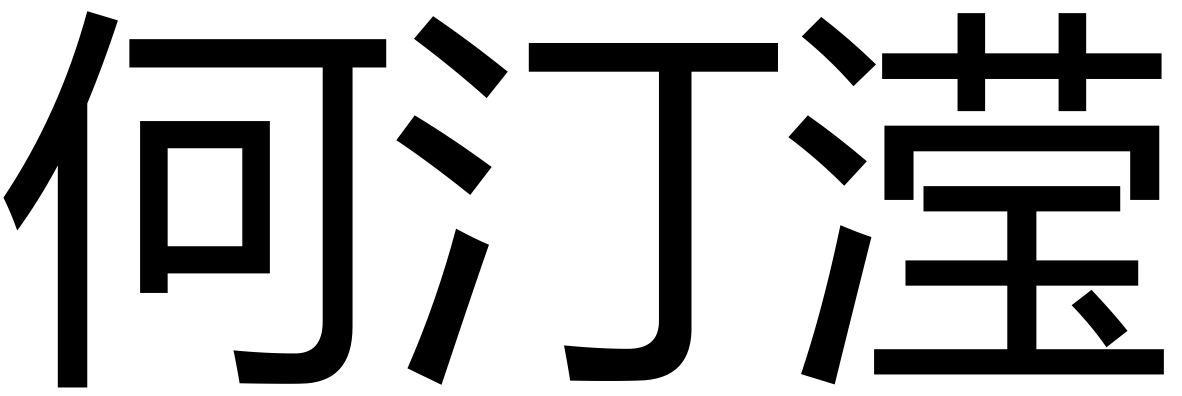}}), M. McCracken, S. Creem-Regehr are with the University of Utah, USA. E-mails: tingying.x.he@gmail.com, \{maggie.mccracken\,$|$\,sarah.creem\}@psych.utah.edu.
    \item D. Hajas is with the Global Disability Innovation Hub, UK. E-mail: d.hajas@ucl.ac.uk.
    \item A. Lex is with Graz University of Technology, Austria and the University of Utah, USA. E-Mail: alex@visdesignlab.net
}

\abstract{%
Visualizations are central to communicating data, yet blind and low-vision (BLV) people often lack support for understanding chart types---knowledge that is essential for interpreting new visualizations and collaborating with sighted peers. 
Prior work found that BLV individuals viewed example tactile charts as more helpful than text-only approaches and preferred them for learning advanced chart types, particularly for understanding spatial layouts and shapes.
Meanwhile, large language models (LLMs) are increasingly used by BLV individuals for chart explanation and question answering (QA), but have been studied primarily for dataset exploration rather than chart-type learning. Existing LLM-based chart QA also shows that users frequently ask about layout and structure, yet struggle with spatial concepts and misdirect questions when mental models are weak. We investigate how LLMs influence chart-type learning and whether tactile learning improves subsequent LLM-supported exploration. We extend our tactile chart learning tools with an LLM chatbot that provides interactive explanations and supports follow-up questions. In an interview study with 12 BLV participants, we compare two learning formats: (1) a tactile chart, a textual explanation, and an LLM chatbot; and (2) a textual explanation and an LLM chatbot. The learning phase was followed by exploration of an unfamiliar dataset using alt text and an LLM. 
\rev{Thematic analysis shows that tactile templates support BLV participants' formation of chart-type mental models, which scaffolds subsequent LLM-mediated data exploration. Text+LLM explanations without tactile support show weaknesses for spatial-reasoning tasks.}

}


\teaser{
  \centering
  \includegraphics[width=\linewidth, alt={A three-panel overview of the modalities used during chart-type learning. Panel (a) shows a screenshot of a web page titled ``Violin Plot Tactile Chart Exploration Instructions,'' organized into steps for orienting to the chart, introducing violin plots, and exploring the tactile model. Panel (b) shows an overhead photograph of a 3D-printed tactile violin plot. Panel (c) shows a screenshot of an LLM chat from an actual participant session. The participant asks whether the upper peaks for Adelie and Chinstrap penguins have the same value and asks for the highest and lowest values for Gentoo; the assistant responds with numerical values and brief explanations.}]{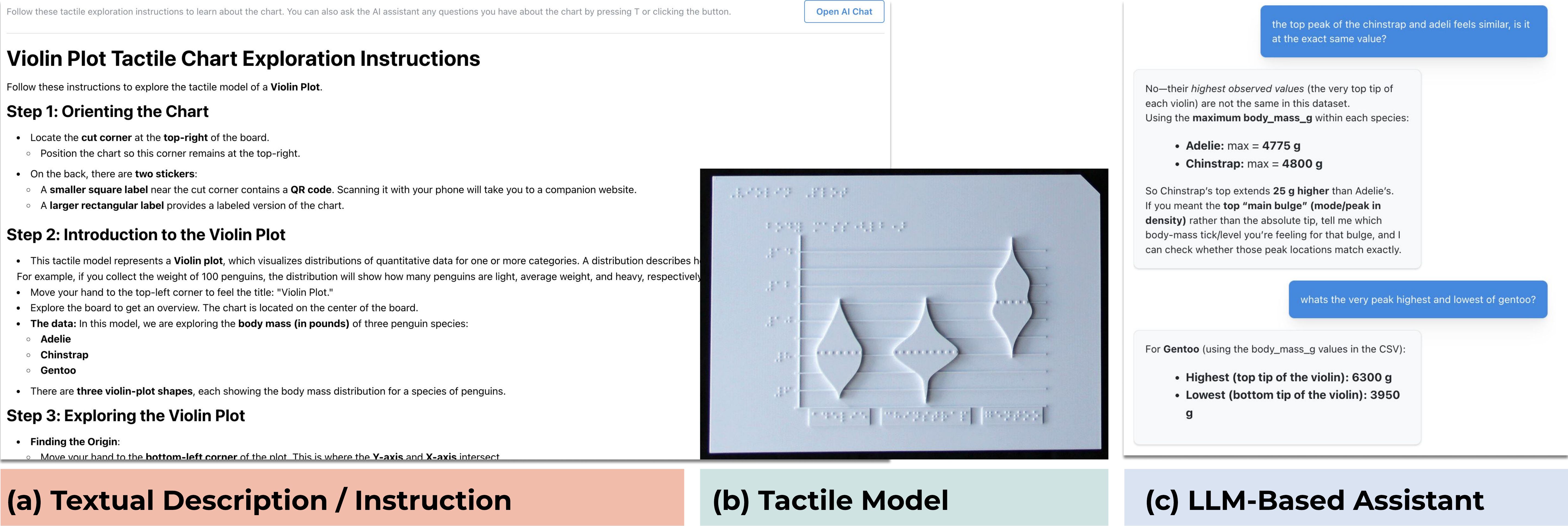}
  \caption{Modalities used to support blind and low-vision participants in learning chart types. We compared \texttt{Tactile+Text+LLM} and \texttt{Text+LLM}. (a) \texttt{Text:} A human-authored description of the chart type.
  In the tactile condition, it guided exploration of the example tactile chart and focused on explaining how the chart works; in the non-tactile condition, tactile-specific content was removed.
  (b) \texttt{Tactile:} A 3D-printed tactile model of a complex chart---a violin plot (shown) or a clustered heatmap---designed to support chart-type learning. (c) \texttt{LLM:} An LLM-based assistant that answered participants’ questions about the chart type, chart, and data; the example shown is from an actual participant session.}
  \label{fig:teaser}
}

\graphicspath{{figs/}{figures/}{pictures/}{images/}{./}} 

\usepackage{tabu}                      
\usepackage{booktabs}                  
\usepackage{lipsum}                    
\usepackage{mwe}                       
\usepackage{ccicons}                   

\usepackage{mathptmx}                  

\usepackage{tabularx}
\usepackage{multirow}

\usepackage{cuted} 

\usepackage[most]{tcolorbox}
\usepackage{xcolor}

\definecolor{insightbg}{HTML}{FFF9CC}
\definecolor{insightborder}{HTML}{E8D94A}

\newtcolorbox{insightbox}{
  colback=insightbg,
  colframe=insightborder,
  boxrule=0.4pt,
  sharp corners,
  left=1pt,
  right=1pt,
  top=1pt,
  bottom=1pt,
  width=\linewidth,
  before skip=4pt,
  after skip=8pt
}
\newcommand{\hquote}[2]{``\textit{#1}''} 

\newcommand{\parhead}[1]{\paragraph{#1}}

\newcommand{\rev}[1]{\textcolor{Crimson}{#1}} 
\renewcommand{\rev}[1]{\textcolor{Black}{#1}} 

\begin{document}



\maketitle

\section{Introduction} 

Visualization is central to scientific communication, decision-making, and public discourse. Accordingly, visualization literacy---the ability to interpret and extract information from charts~\cite{Lee:2017:VLAT}---is increasingly important for full participation in professional and civic life.  
People who are blind or have low vision (BLV) need to \textit{read} visualizations~\cite{Lee:2020:Reaching, Siu:2021:Covid, Joyner:2022:Visualization}, \textit{collaborate with co-workers in analyzing} them, and \textit{produce} visualizations themselves~\cite{Zong:2024:Umwelt}. \rev{This need is prevalent: approximately 7--8 million Americans live with vision impairment or blindness, and globally, an estimated 36 million people are blind, while 217 million have moderate-to-severe vision impairment~\cite{Ackland:2017:World}.} Yet BLV users continue to face persistent access barriers when charts are delivered primarily through visual formats \cite{Lee:2020:Reaching, Siu:2021:Covid, Joyner:2022:Visualization}. 
In practice, many BLV users rely on alternative text (alt text) to access charts. However, most alt text is effective only when readers already know the chart types and understand how the encodings work~\cite{Joyner:2022:Visualization}. Shared chart-type understanding is also critical in blind-sighted collaboration, where teams might refer to the chart features during their communication.

\rev{Large language models (LLMs) and conversational AI systems are increasingly being explored as interfaces for supporting BLV individuals in accessing data and exploring charts~\cite{Seo:2024:MAIDRAI, Gorniak:2024:VizAbility,Das:2026:Making,Hohenwalde:2025:Enhancing}. Recent accessibility research suggests that conversational support can help BLV users understand specific visualizations, but it also shows that chart-type knowledge remains important~\cite{Kim:2023:Exploring, Seo:2024:MAIDRAI, Gorniak:2024:VizAbility}.}
\rev{For example, prior work has observed that, when using chart question-answering systems, BLV users frequently ask about chart layout, try to understand unfamiliar chart types, and experience greater cognitive load when working with unfamiliar visualizations.} These findings suggest that LLM-based chart exploration may be most useful when BLV users have a mental model of the chart type.

Although BLV participants commonly report being familiar with basic charts, such as bar, line, and pie charts, they report much lower familiarity with more advanced or complex chart types, such as area charts or violin plots \cite{Engel:2018:User, Engel:2017:Improve, Wang:2022:Seeing}. A few studies have begun to investigate how to better support BLV individuals' understanding of advanced chart types, mainly focusing on text or tactile charts~\cite{McNutt:2025:Accessible, Kim:2023:Explain, He:2026:Using}.
Our own prior work found that tactile example charts, particularly when paired with structured exploration instructions, can help BLV people build transferable knowledge of complex chart types \cite{He:2026:Using}. These tactile charts are especially valuable for supporting BLV individuals in understanding the spatial layout and shape of chart elements that are difficult to communicate through text alone.

\rev{At the same time, LLMs offer a promising complement to tactile charts, because they can provide interactive, on-demand explanations through a digital interface. While the digital and flexible nature of LLMs makes them an invaluable resource for BLV users to understand specific datasets, it is less clear whether they are beneficial for learning about new chart types in the first place.
Prior work on visualization education has explored LLMs for teaching chart types and improving visualization literacy with sighted users~\cite{Choe:2025:Enhancing, Wang:2025:VizTA}.
Prior multimodal accessibility work has also established that hybrid techniques combining tactile with speech or sonification reduce workload and improve scalability for complex tasks~\cite{Yu:2003:Evaluation}. 
Therefore, in this work, we combine tactile charts with exploration instructions and an LLM-based assistant to support BLV individuals in learning advanced chart types.
Our goal is to understand the utility of tactile models for chart learning in the context of large language models: whether tactile models still provide unique benefits when LLM explanations are available, whether they are supplanted by LLM-generated textual explanations, or whether their value lies in complementing LLMs by grounding textual explanations in spatial experience. We also examine how chart-type learning shapes participants' subsequent exploration of new datasets in the same chart type with LLM support.}

\rev{To investigate this question, we conducted an exploratory mixed-methods study with 12 BLV participants.
We selected two complex chart types commonly used in scientific contexts, \textbf{clustered heatmaps} and \textbf{violin plots}. 
Participants learned two chart types, each under one of two conditions: \textbf{(1) tactile model + textual instruction + LLM-based assistant} or \textbf{(2) textual instruction + LLM-based assistant}.
We used 3D-printed tactile example charts that we developed previously~\cite{He:2026:Using}, which BLV participants had found to be well designed and preferable for supporting their chart learning, relative to human-authored textual descriptions.}
We also developed and iteratively refined a study website in collaboration with our blind co-author, where participants could read instructions and alt text and interact with an LLM-based assistant.

In each condition, participants first learned the chart type by reading the instructions, exploring the tactile chart when available, and asking the LLM assistant questions. The tactile charts were intended to support chart-type learning rather than communicate a specific dataset.
Participants then explored a new dataset in the same chart type using a manually curated alt-text description and the LLM assistant. 
We analyzed queries, learning outcomes, and participants' thoughts and preferences on the helpfulness of different chart-type learning methods. We found that learning with tactile charts is perceived as helpful for subsequent data exploration, and participants prefer the combination of the three modalities (tactile chart + instructions + LLM).

The main contribution of this paper is the interview results from 12 BLV participants who learned and explored advanced charts using tactile charts and LLMs. These results show the perceived value of chart-type learning with tactile charts for supporting later exploration of new datasets with LLMs, reveal the benefits and limitations of LLM support in learning about new chart types and datasets, and highlight participants’ preferences across learning modalities. 
We also provide a collection of LLM queries from BLV individuals during the learning and exploration of complex charts, along with insights into the patterns these queries reveal. Our results are useful in an educational context for BLV individuals, highlighting that to teach about data visualizations, LLMs can supplement, but not replace means of communicating spatial data, such as tactile charts. 

\section{Related Work}
In this section, we review prior work on the accessibility of complex visualizations, chart-type learning and comprehension for BLV individuals, and the use of LLMs to support visualization accessibility.

\subsection{Visualization Accessibility of Advanced Chart Types}

Visualizations are a powerful medium for exploring data and communicating insights, yet their dependence on vision creates barriers for BLV individuals \cite{Choi:2019:Visualizing}. To address these barriers, researchers have examined what information BLV readers seek from charts~\cite{Lundgard:2022:Accessible}, developed a range of sensory substitutions---including speech \cite{Zong:2022:Rich, McNutt:2025:Accessible, Lundgard:2022:Accessible}, haptic or tactile representations~\cite{Yang:2020:Tactile, Engel:2019:User, Goncu:2011:GraVVITAS}, sonification~\cite{Hoque:2023:Accessible, Daunys:2008:Sonification, Franklin:2003:Pie}---that translate data and visual structure of charts into non-visual modalities, and proposed authoring and evaluation guidelines for these modalities. 
However, most visualization accessibility research has historically focused on basic chart types (e.g., bar, line, pie charts) ~\cite{Kim:2021:Accessible, Wimer:2024:Beyond}.

Providing equal access to complex visualizations is a significant challenge, and research on making advanced chart forms more accessible is still under-represented  ~\cite{Marriott:2021:Inclusive}, but is emerging, recognizing the need to make complex charts accessible for BLV users in an educational and professional context. Recent tools that make visualization specifications navigable via screen readers, such as Olli \cite{Blanco:2022:Olli, zong:2025:semantic} or Data Navigator~\cite{Elavsky:2024:Data}, have incorporated more complex designs (e.g., multi-series and faceted charts, stacked charts, or maps). Multimodal data representation techniques (e.g., MAIDR~\cite{Seo:2024:MAIDR}) are also expanding to advanced chart types (e.g., heatmaps). Beyond general-purpose systems, some individual advanced chart families have received attention, such as automated alt-text approaches for UpSet plots~\cite{McNutt:2025:Accessible} and genomics visualizations \cite{Smits:2024:AltGosling}, as well as work on tactile representations for network visualizations (e.g., TADA~\cite{Zhao:2024:TADA}) and accessibility of geospatial visualizations~\cite{Li:2024:AltGeoViz}. 
This work further contributes to making complex visualizations accessible by evaluating chart-type learning methods.

\subsection{Chart-Type Understanding for BLV Individuals}

Visualization literacy refers to the ability to read, interpret, and extract insights from visualizations \cite{Lee:2017:VLAT}, as well as create visualizations~\cite{ge:2025:avec}. Prior work has identified barriers to visualization literacy by examining the mental models that underlie incorrect chart interpretations \cite{Nobre:2024:Reading}. To better support BLV individuals in reducing these barriers, Zong et al.\ argue that researchers need to avoid visual-focused approaches, and instead further develop effective non-visual channels for data exploration \cite{Zong:2024:Umwelt}. However, many current accessibility approaches still depend heavily on visual representations, which makes chart-type understanding especially important. For example, prior studies show that chart-type understanding is essential for interpreting alt text \cite{Jung:2022:Communicating}, and that BLV users frequently ask chart-type questions when interacting with chart question-answering systems \cite{Kim:2023:Exploring}. In addition, many professional tasks require blind-sighted collaboration. Therefore, chart-type understanding remains a critical skill for BLV individuals' full participation in professional and everyday contexts.

While BLV participants report familiarity with basic charts (e.g., bar, line, or pie charts), familiarity is much lower for more advanced or complex chart types (e.g., stacked bars, area charts, donut charts, violin plots) \cite{Engel:2018:User, Engel:2017:Improve, Wang:2022:Seeing}. To bridge this gap, a few works study how to support BLV individuals in learning unfamiliar chart types. He and Yu~\cite{He:2024:Charting} adopt multisensory learning strategies to support visualization literacy education for BLV children learning bar, line, and scatter plots. Kim et al.~\cite{Kim:2023:Explain} propose an automatically generated explanation system spanning 50 chart specifications and compare pedagogical strategies (i.e., linking to familiar chart types vs.\ not, declarative vs.\ procedural framing, abstract vs.\ concrete descriptions) for explaining charts to BLV people. Smits et al.~\cite{Smits:2024:Explaining} investigate staged explanation to help BLV individuals learn unfamiliar genomics visualizations. Mei et al.~\cite{Mei:2025:Benthic} introduce Benthic, a hypergraph-based approach that represents charts and diagrams in perceptually congruent structures for screen reader users. While not framed as instruction, it supports users in developing a clearer structural understanding of how a chart is represented. Finally, in our previous work, we studied the use of example tactile charts as learning materials for complex visualizations and reported that tactile examples paired with exploration instructions are a preferred and effective learning method compared to text-only explanations ~\cite{He:2026:Using}. We found tactile models especially helpful in supporting BLV individuals to understand chart layout and the shape of chart elements; therefore, we continue to adopt this method in this study. 

Researchers have explored LLMs to improve visualization literacy and help novice users understand unfamiliar chart types in the general population~\cite{Choe:2025:Enhancing, Wang:2025:VizTA}. However, to our knowledge, prior work has not specifically examined how LLMs can support BLV individuals in learning unfamiliar chart types. Our work addresses this gap by integrating LLM-based assistance into a BLV chart-learning workflow.

\subsection{LLMs for Visualization Accessibility}

Generative AI, particularly LLMs such as ChatGPT, has quickly become a common interface for information access, and BLV individuals are often eager to adopt these tools for everyday tasks. Recent HCI and accessibility studies have examined how BLV users use GenAI tools~\cite{Tang:2025:Everyday, Adnin:2024:Look}, including preferences and challenges. Within visualization, one line of work uses LLMs to produce textual descriptions of charts. Earlier approaches to automated chart description largely relied on templates and rules~\cite{Demir:2010:Interactive,Moraes:2014:Evaluating,Demir:2008:Generating}, and a few recent works have begun using LLMs, which offer greater expressive flexibility \cite{Yan:2025:Chart}. Examples include DATATALES by Sultanum and Srinivasan~\cite{Sultanum:2023:DATATALES}, which supports authoring textual narratives grounded in a given visualization, and AutoVizuA11y by Duarte et al.~\cite{Duarte:2024:AutoVizuA11y}, which uses an LLM to automatically generate descriptions of web-based visualizations and provide statistical insights derived from the underlying dataset. At the same time, McNutt et al.~\cite{McNutt:2025:Accessible}
found that LLM-generated textual descriptions are, in some cases, inferior to rules-based generation, especially when considering hallucinations.

Another line of work uses LLMs to support conversational chart question answering, enabling BLV users to query visualizations and iteratively explore data through dialogue. Kim et al.~\cite{Kim:2023:Exploring} provide foundational evidence for this direction through a Wizard-of-Oz study that analyzed BLV participants' queries about four visualizations. Seo et al.~\cite{Seo:2024:MAIDRAI} integrate an LLM into a multimodal data representation framework and examine how BLV individuals interact with multimodal LLM assistance for chart exploration. Gorniak et al.~\cite{Gorniak:2024:VizAbility} present VizAbility, an LLM-based conversational system for exploring common visualization types, including bar charts, line charts, scatterplots, and choropleth maps.
Across these studies, researchers observed that BLV users frequently seek chart-type understanding and experience greater cognitive load when interpreting LLM responses about unfamiliar charts. Participants also frequently ask questions about the chart's spatial layout. These findings suggest that users may benefit from becoming familiar with chart types in advance and that this knowledge is transferable between specific datasets. 

\begin{table*}[t]
\centering
\small
\caption{Participant demographics. IP = degree in progress; $\bar{X}$ is the mean; $Y$ is the mode. Note that P6 and P8 were assigned a participant ID but did not end up participating in the study and are hence absent from the table.}
\resizebox{\textwidth}{!}{%
\begin{tabular}{ccccccccc}
\toprule
\textbf{pID} & \textbf{Age} & \textbf{First Interview} & \textbf{Data Intensive} & \textbf{Braille} & \textbf{Vision Loss} & \textbf{Onset of} & \textbf{Highest Level} & \textbf{AI Tools}\\ 
   & \textbf{(Gender)} & \textbf{Chart (modality)} & \textbf{Career} & \textbf{Years Exp.} & \textbf{Level} & \textbf{Vision Loss} & \textbf{of Education}  &  \textbf{Used} \\ 
 \toprule
P1 & 29 (M) & Heatmap (tactile) & No & 25 & No residual vision & Blind since birth & Bachelor's & Gemini \\ 
P2 & 48 (M) & Violin (text) & Yes & 7 & No residual vision & Blind since birth & Master's & BeMyAI, ChatGPT, Gemini \\ 
P3 & 32 (M) & Violin (tactile) & Yes & 15 & Peripheral vision loss & Blind since birth & Associate's (IP) & BeMyAI \\ 
P4 & 40 (M) & Heatmap (text) & Yes & 34 & No residual vision & Blind since birth & Not disclosed & ChatGPT, Gemini \\ 
P5 & 34 (F) & Heatmap (tactile) & No & 26 & No residual vision & Blind since birth & Bachelor's (IP) & ChatGPT, Gemini \\ 
P7 & 23 (M) & Violin (tactile) & No & 19 & No residual vision & Blind since birth & Bachelor's & ChatGPT, Claude, Gemini \\ 
P9 & 44 (F) & Heatmap (tactile) & No & 42 & Light perception & Lost vision suddenly & Master's & BeMyAI, Gemini \\ 
P10 & 40 (F) & Violin (tactile) & No & 35 & No residual vision & Blind since birth & Bachelor's & Aira, ChatGPT, Copilot\\ 
P11 & 34 (F) & Violin (text) & Yes & 26 & No residual vision & Lost vision suddenly & Master's & ChatGPT, Gemini\\ 
P12 & 40 (M) & Heatmap (text) & Yes & 30 & No residual vision & Lost vision gradually & Master's & Gemini \\ 
P13 & 29 (M) & Violin (text) & Yes & 25 & Light perception & Blind since birth & Bachelor's & ChatGPT, Claude\\ 
P14 & 63 (F) & Heatmap (text) & Yes & 55 & Severe low vision & Blind since birth & Master's & ChatGPT, Gemini \\ 

\midrule
& $\bar{X}=$  
&
& 7/12  
& $\bar{X}=$ 
& $Y$ = 
& $Y$ =
& $Y$ =
& $Y$ =
\\
& $38 \pm 11$  
&
& Yes  
& $28 \pm 13$ 
& No residual vision (8)
& Blind since birth (9)
& Master's (5) 
& Gemini (9)
\\
\bottomrule
\end{tabular}
}
\label{tab:demographics}
\end{table*}


\section{Study Stimuli}

To prepare for our evaluation of the utility of LLMs and tactile charts for teaching chart types to BLV individuals, we developed the study stimuli in close collaboration with our blind co-author. We used two stimuli: a study website with an LLM-based chatbot, and a set of tactile charts. We describe these components in detail below.

\subsection{Tactile Charts}

We selected two chart types for our evaluation: a clustered heatmap and a violin plot, both advanced chart types commonly used in scientific reports. 
We designed the tactile chart models for these two chart types in our previous work~\cite{He:2026:Using}.
These tactile models are example charts designed specifically for chart-type learning. We developed them in close collaboration with two BLV researchers and followed established best practices for tactile graphics. We also designed accompanying exploration instructions to guide learners’ use of the models. These materials can all be accessed on our website \href{https://vdl.sci.utah.edu/tactile-charts/}{vdl.sci.utah.edu/tactile-charts/}. In our previous study, participants responded positively to the design of these materials, and tactile charts paired with exploration instructions were described as a preferred and helpful way for BLV individuals to learn these complex chart types. An example of the violin-plot tactile chart is shown in Figure~\ref{fig:teaser}. 

For this study, we thus followed the same logistical procedure as before. We 3D-printed a new set of tactile charts and attached the labels from the sighted version, as well as a QR code linking to the corresponding exploration instructions, to the back of each model.

\subsection{Study Website and the LLM-based Chatbot}
We implemented our online study website for the two study phases: \textit{chart learning} with a simple dataset and \textit{testing} with a more complex dataset that is not available as a tactile chart. The website is implemented with the reVISit framework~\cite{Cutler:2026:Revisit2}. The website is available at  \href{https://vdl.sci.utah.edu/tactile-and-LLM-revisit/}{vdl.sci.utah.edu/tactile-and-LLM-revisit/}; the code is open source and available on \href{https://github.com/visdesignlab/tactile-and-LLM-revisit/}{github.com/visdesignlab/tactile-and-LLM-revisit/}. To support data collection for the evaluation study, the website also records audio, user interactions, user prompts, and chatbot responses.

For the chart-type learning phase, the website presents pre-authored instructions tailored either for use with the tactile chart or for text-only learning, depending on the condition (\autoref{fig:teaser}(a)). For the testing phase, the website presents pre-authored alt text for a new dataset using the chart type the participant has just learned. 
At each step of the study (each page of the website), participants can open a modal window to chat with an assistant, which is an LLM-based chatbot tailored to the current study state (\autoref{fig:teaser}(c)). 
We use the OpenAI Responses API and the GPT-5.2 model. Below, we describe the implementation details of this chatbot.

\parhead{Access to Chart-Specific Materials}
The chatbot can access three types of chart-specific resources: (1) the chart image, (2) the underlying dataset, and (3) the learning materials. Chart images are stored as pre-uploaded PNG files in the OpenAI file system,
which the model can consume as an image and leverage GPT-5.2’s vision capability to interpret visual details when needed. The underlying dataset is loaded from a CSV file. Learning materials are provided as Markdown.

During the chart-type learning phase, the chatbot may request all three resources (image, dataset, and instructions). During the new-dataset exploration phase, the chatbot can request the chart image and the dataset. 
The specific image, dataset, and (when applicable) instructions returned are determined by the current study condition (i.e., chart type, learning modality, and study phase). To keep responses fast and to avoid unnecessary data transfer, we provide these resources only on demand via Responses API tool calls.

\parhead{System Prompt}
The chatbot’s behavior was controlled by a system prompt selected based on: (1) study phase (chart-type learning vs.\ new-dataset exploration), (2) learning modality during the learning phase (with vs.\ without the tactile chart), and (3) the current chart type. The selected template is then parameterized with chart-type–specific context so the assistant’s guidance matches what the participant is currently learning or exploring.

The system prompt begins by defining the assistant’s role and goal in the study. During the learning phase, it explicitly states which materials the participant has access to, so the assistant’s guidance remains aligned with the assigned condition. Across all conditions, the prompt specifies which chart-specific artifacts the chatbot can access and enforces a gating policy: the assistant should answer from general knowledge by default and request chart-specific artifacts only when the participant explicitly asks for them or when the question clearly requires exact values or visual details; otherwise, it asks a brief clarifying question. The prompt also constrains response style (e.g., clarity and concision) to keep the interaction consistent and easy to follow.

\parhead{Maintaining Conversation State}
We maintain continuity across turns using the Responses API’s conversation-state mechanism. It keeps context from earlier turns, and thus allows the model to continue a multi-turn conversation without resending the full history.

\parhead{Accessibility for Screen-Reader Users} To make sure our study website is easy to use with screen readers, we tested and iterated with our blind co-author. Our final prototype has the following dedicated features: First, the interface uses visually hidden live regions to announce interaction state changes (e.g., ``AI is thinking'') and to read out newly available assistant responses as they arrive. Second, to support efficient navigation of the transcript, we mark user messages with Heading 5 and assistant messages with Heading 6. These consistent structural landmarks enable quick navigation by headings without disrupting the page’s primary heading hierarchy, because H5/H6 are typically unused in interfaces. Third, each message is preceded by a screen-reader-only role label (i.e., ``You said'' and ``Assistant said'') to clearly identify the speaker. Finally, we label all interactive controls with aria-labels and manage focus and scroll behavior so keyboard and screen-reader users can keep their place as new content appears.

\section{Evaluation Study}
With our study website and tactile charts, we conducted interviews with 12 BLV participants to evaluate the use of tactile charts and an LLM for learning unfamiliar chart types and their impact on subsequent exploration of new datasets.  The study was reviewed by the University of Utah IRB and approved as exempt from full-board review (\textnumero\,IRB\_00180924).

\subsection{Study Design}

Our study design and methodology followed our prior work~\cite{He:2026:Using, McNutt:2025:Accessible}, which used a similar target group. We reused our established protocol of interviewing and providing tactile models~\cite{He:2026:Using}, but restate our process here for convenience. We employed a mixed-design method with two factors: (1) chart type: violin plot or clustered heatmap, and (2) learning modality: a tactile chart with exploration instructions and an LLM-based assistant (\texttt{Tactile+Text+LLM}), or a textual explanation and an LLM-based assistant (\texttt{Text+LLM}). Each participant learned both chart types, with each chart type using a different learning modality. 
\rev{To minimize order effects, we counterbalanced across the four possible order combinations (2 chart-type orders × 2 learning-modality orders).}
We adopted this approach because we considered the expected interview length and recruitment challenges, while still allowing participants to experience both chart types and both learning modalities, thereby enabling comparison across modalities. We conducted semi-structured interviews led by two researchers ~\cite{akbaba2023two}. The average length of interviews was 135 $\pm$ 25 min (SD). We compensated each participant with a \$100 Amazon gift card.

\subsection{Study Material Preparation}
\label{sec:study-material-preparation}

We adopted the datasets, tactile charts, instructions, and alt text from our previous work~\cite{He:2026:Using}. Participants encountered two datasets for each chart type: a simple dataset for learning and a more complex dataset for assessing learning outcomes. The datasets used in the violin plot condition were body mass of penguins (simple)~\cite{Horst:2020:PalmerPenguins} and the Human Development Index across five continents (complex). The datasets used in the clustered heatmap condition were movie genres and actors (simple) and agreements on sociocultural topics in European countries (complex)~\cite{EVS}. 

For the simple datasets, we created the tactile charts (see \autoref{sec:tactile-charts}  \rev{or \href{https://vdl.sci.utah.edu/tactile-charts/}{vdl.sci.utah.edu/tactile-charts/}}), wrote the corresponding instructions, and authored alt text (see \autoref{sec:simple-alt-text}). For the complex datasets, we created the visualizations (see \autoref{sec:complex-charts}) and authored alt text (see \autoref{sec:complex-alt-text}). Following established guidelines~\cite{Lundgard:2022:Accessible, McNutt:2025:Accessible}, the alt text included a short high-level summary followed by more detailed information, such as statistical details and trend analysis. In this study, because we wanted participants to have open questions for exploration with the LLM assistant, we retained only the high-level summary.

\rev{All textual materials and the LLM-based chatbot are available on the study website: \href{https://vdl.sci.utah.edu/tactile-and-LLM-revisit/}{vdl.sci.utah.edu/tactile-and-LLM-revisit/}.}
We also prepared the interview questions, including questions about participants’ backgrounds, their understanding of the charts and datasets, and their comparisons of the learning modalities (see \autoref{sec:interview-questions}).

\subsection{Participants}

We recruited participants by posting in relevant social media groups. To be eligible, participants had to be at least 18 years old, speak English, be legally blind (best-corrected visual acuity of 20/200 or worse, or a visual field  $\leq$20$^\circ$ in the better-seeing eye), read Braille, and use a screen reader on their computer. We further required participants to be working or enrolled in school at least part-time, because our target context was professional data analysis. Participants also needed to have a U.S. mailing address to receive the tactile chart.

We first recruited one eligible participant who had taken part in our previous study and still had the tactile chart in hand for a pilot study. 
The pilot allowed us to verify the procedure, estimate session duration, and confirm the accessibility and usability of the study website.

We recruited 12 new eligible participants for the formal study. 
One participant (P6) dropped out on short notice due to a health issue. One participant (P8) attempted to participate but could not due to technical issues with Zoom, which we were unable to resolve within the allotted time. To maintain counterbalancing across conditions, we recruited two additional eligible participants. Ultimately, 12 participants completed the formal interview study, and their results were included in the following analysis. Both the pilot participant and P8, who attempted to participate, received the same compensation as other participants. \autoref{tab:demographics} summarizes participants' demographics, and full participant background information is provided in the supplementary materials.

\subsection{Procedure}

\begin{figure*}
    \centering
    \includegraphics[width=1\linewidth, alt={A horizontal flow diagram of the study procedure. It begins with background questions, followed by two study parts with different chart-type and learning-modality pairings. Part 1, illustrated as Tactile+Text+LLM with a violin plot, proceeds through chart-type instruction, questions about understanding, simple alt text with the LLM, complex alt text with the LLM, and questions about understanding and factual observations. Part 2 repeats the same sequence, illustrated as Text+LLM with a clustered heatmap. The procedure ends with comparative questions about the two learning modalities. Arrows connect all stages from left to right.}]{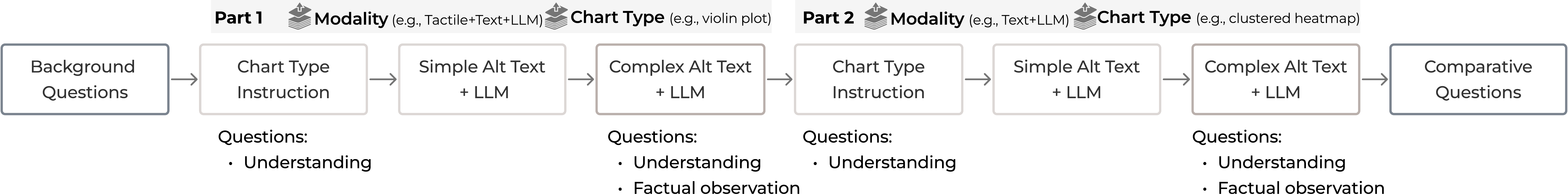}
    \caption{\rev{Study procedure. Participants first answered background questions, then completed two parts with different modality/chart type combinations. In each part, they learned the chart type with or without a tactile chart, practiced exploring a simple dataset with alt text and the LLM assistant, and then explored a complex dataset. Participants answered questions after chart-type learning and complex-dataset exploration. The study ended with comparative questions.}}
    \label{fig:study-design}
\end{figure*}

\rev{In this section, we present the study procedure for each participant. Depending on the condition and phase, participants completed different activities: they read textual materials, including instructions and alt text; used the LLM-based chatbot on the study website; touched the tactile charts; and answered the study questions. \autoref{fig:study-design} shows the study procedure. We also describe all study materials in \autoref{sec:study-material-preparation} and provide the full set of study questions in \autoref{sec:interview-questions}.}

\parhead{Pre-Interview} 
We screened interested participants for eligibility via chat or email. For eligible participants, we shared the study procedure and consent process and scheduled a 2-hour online interview. Participants then reviewed and signed the consent form in REDCap. 
Based on each participant’s assigned condition, we sent the corresponding tactile model via mail \cite{deGreef:2021:Interdependent} and provided participants with links to both the tactile-model instructions and the textual explanations for the two chart types ahead of time. We included these links both as QR codes in the package and in a follow-up email. Participants could explore the tactile model and instructions before the session, but we did not require this preparation.

\parhead{Background Questions (13 $\pm$ 2 min)} After participants agreed to audio and screen recording, we asked background questions about demographics, vision loss, prior experience with screen readers, Braille, tactile graphics, AI assistants, and familiarity with clustered heatmaps and violin plots. We then sent each participant a link to the study website via Zoom chat or email and asked them to share their screen.

In the rest of the session, we guided participants through the website page by page. Participants could take as much time as needed to read the current page and chat with the assistant. To keep the procedure consistent across participants, we did not allow participants to return to previous pages. We also instructed participants on the chatbot's functionality to ensure they understood how to use it during the study.

\parhead{Chart Type Instructions (18 $\pm$ 6 min)} Participants first read an instruction page about the chart type. Depending on the assigned modality condition, they explored the tactile chart or learned from the text-only materials. Participants could open the chatbot and chat with the assistant at any time. We then asked five questions to assess chart-type understanding: three were shared across both chart types (e.g., ``What types of data are best suited for visualization using a clustered heatmap / violin plot?''), and two were chart-specific (e.g., ``If a violin’s shape has three peaks, what does that imply about the underlying data?'').

\parhead{Simple Alt Text (14 $\pm$ 3 min)} Next, participants read the alt text of the chart used in the instruction page. This step served as practice for using the LLM chatbot to explore a dataset in the chart type they had just learned. We encouraged participants to chat with the AI assistant as much as possible to learn more about the dataset, and we provided three example queries as inspiration. These example queries were adapted from Seo et al.'s study ~\cite{Seo:2024:MAIDRAI} and covered information in Levels 2 to 4 from Lundgard et al.'s semantic levels~\cite{Lundgard:2022:Accessible}, as Level 1 information is already mostly covered in the alt text. We asked all participants to read the example queries, but we did not require them to ask those questions.

\begin{table}[t]
\centering
\small
\caption{Rates of answer quality for chart understanding, new dataset understanding, and factual observations about the new dataset under the two learning modality conditions (N = 12).}
\begin{tabularx}{\linewidth}{>{\raggedright\arraybackslash}X l c c c}
\toprule
  & \textbf{Training} & \textbf{Correct} & \textbf{Partially Correct} & \textbf{Wrong} \\
\midrule
\multirow{2}{\linewidth}{Chart type understanding} 
& \textbf{Tactile+Text+LLM} & 30.00\% & 36.67\% & 33.33\% \\
& \textbf{Text+LLM}    & 31.67\% & 45.00\% & 23.33\% \\
\midrule
\multirow{2}{\linewidth}{New dataset understanding} 
& \textbf{Tactile+Text+LLM} & 62.50\% & 25.00\% & 12.50\% \\
& \textbf{Text+LLM}   & 62.50\% & 25.00\% & 12.50\% \\
\midrule
  &  & \textbf{Good} & \textbf{OK} & \textbf{Bad} \\
\midrule
\multirow{2}{\linewidth}{New dataset factual obs.} 
& \textbf{Tactile+Text+LLM} & 50.00\% & 27.78\% & 22.22\% \\
& \textbf{Text+LLM}    & 44.44\% & 25.00\% & 30.56\% \\
\bottomrule
\end{tabularx}
\label{tab:factual-questions-correctness}
\end{table}

\parhead{Complex Alt Text (23 $\pm$ 7 min)} Next, participants read the alt text for a new, more complex dataset in the same chart type and used the LLM assistant to explore it. We asked them to imagine encountering this chart in everyday life and trying to understand it as fully as possible. We encouraged them to use the LLM for in-depth exploration. Afterward, they answered five questions assessing their understanding (e.g., ``Imagine you have a friend who has never encountered this chart before. How would you describe this chart to them?''). Participants could take as much time as they needed for the exploration, but they were not allowed to read the alt text or use the LLM while answering these questions. 
We then asked participants to rate their understanding of the dataset, describe any difficulties they encountered, reflect on whether the assistant helped, and share any additional feedback about their experience with this chart type.

\parhead{Comparative Questions (12 $\pm$ 5 min)} Participants completed the procedure for both chart types. We then asked them to compare the two training formats (\texttt{Tactile+Text+LLM} vs.\ \texttt{Text+LLM}), state their preference, and explain their reasoning. We also asked participants to rate and justify: (1) the helpfulness of the tactile charts for chart-type learning, (2) whether tactile learning affected their subsequent new-dataset exploration, (3) the helpfulness of the LLM during learning and exploration, and (4) the perceived usefulness of each training format for BLV education. Finally, we invited any additional feedback.

\subsection{Quantitative Results}

In this section, we report our quantitative results, including our assessments of participants’ responses to questions about the chart types and datasets, as well as their subjective ratings of the learning modalities.

\paragraph{Understanding and Factual Questions}
Two researchers independently coded all participant responses and resolved disagreements through discussion.
For questions asked after the instructions that address chart type understanding, we coded responses as correct, partially correct, or incorrect. For questions asked after new dataset exploration, responses about the overall understanding of the dataset were coded using the same categories. For open-ended questions about factual observations in the new datasets, such as numerical characteristics, trends, and interesting or surprising findings, responses were coded as good, OK, or bad.
As shown in \autoref{tab:factual-questions-correctness}, \texttt{Tactile+Text+LLM} training did not improve performance on understanding questions, consistent with our prior results comparing pre-authored alt text and tactile charts~\cite{He:2026:Using}. For factual observations about the new datasets, responses in the \texttt{Tactile+Text+LLM} condition were slightly higher in quality (50.00\% good vs.\ 44.44\% in the \texttt{Text+LLM} condition).


\paragraph{Subjective Ratings}

\begin{figure}[t]
    \centering
        \includegraphics[width=1\linewidth, alt={A series of diagrams shows participant ratings and preferences. Panel (a) shows histograms of ratings for understanding complex datasets, split by chart type (violin plot and clustered heatmap) and learning modality (Tactile+Text+LLM and Text+LLM). Average ratings range from 3.83 to 4.33. Panel (b) shows the preferred learning modality, with 11 participants choosing Tactile+Text+LLM and one indicating that their preference depends on the chart type. Panel (c) shows ratings for the usefulness of the two learning modalities in BLV education, with Tactile+Text+LLM rated higher than Text+LLM. Panel (d) shows ratings for the helpfulness of tactile charts in understanding chart types and subsequently exploring new datasets with the LLM, with most ratings at 4 or 5. Panel (e) shows ratings for the helpfulness of the LLM chatbot, which was rated more helpful for exploring new datasets than for understanding chart types.}]{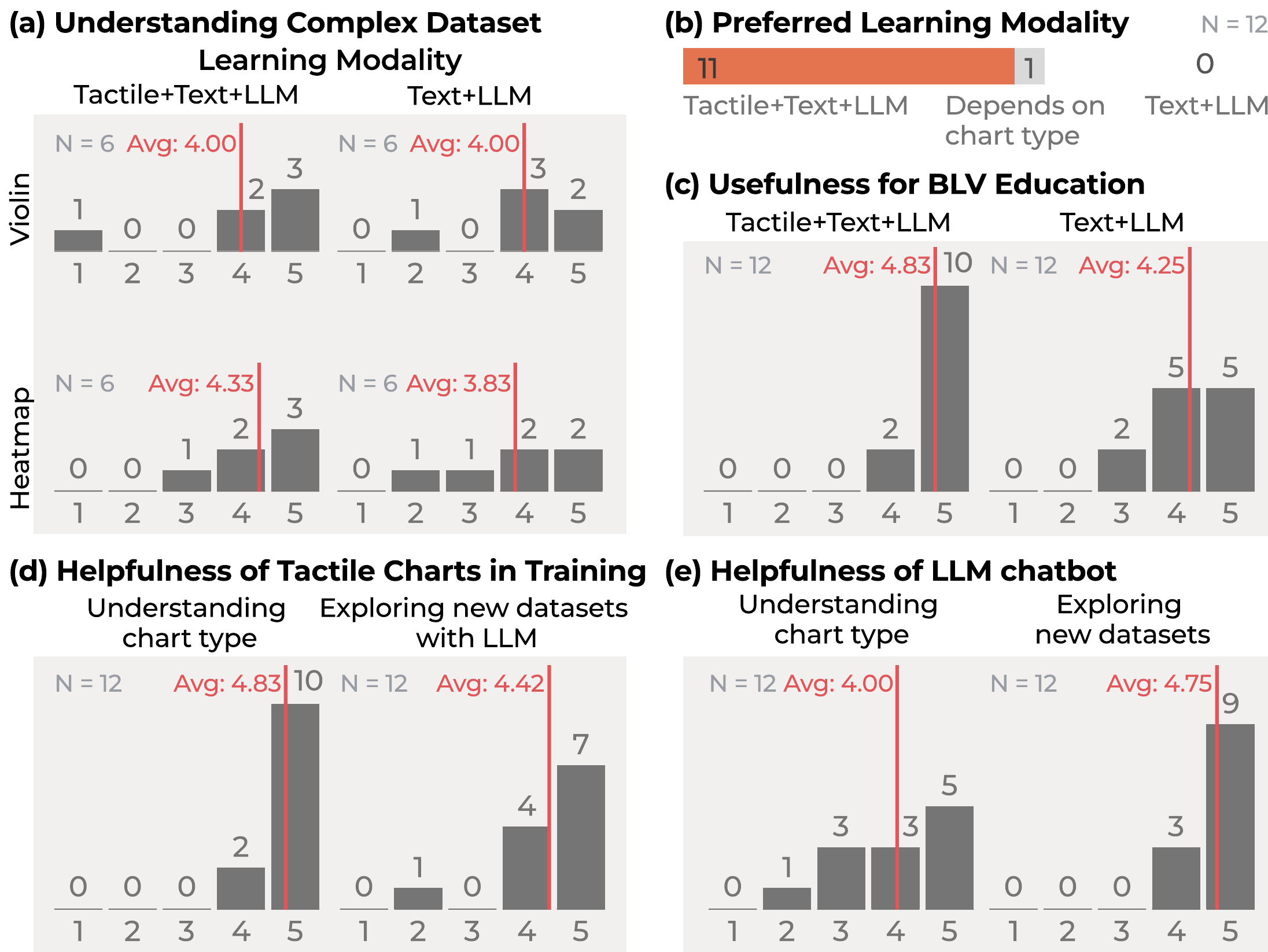}
    \caption{\textbf{Participants prefer to learn chart types when all modalities (tactile models, LLM assistants, and text descriptions) are available.}  (a), (c), (d), and (e): Distributions of participants’ responses to different questions on a 5-point Likert scale (1 = strongly disagree, 5 = strongly agree). (b): Participants’ preferred learning modality.}
    \label{fig:ratings}
\end{figure}

\autoref{fig:ratings} summarizes participants’ subjective ratings of the two learning modalities. 
Overall, perceived understanding of the new complex datasets was similar across conditions (\autoref{fig:ratings}(a)). For the violin plot, both conditions received the same mean rating ($M = 4.00$). For the clustered heatmap, participants reported higher perceived understanding after \texttt{Tactile+Text+LLM} training ($M = 4.33$) than after \texttt{Text+LLM} training ($M = 3.83$). 
Participants strongly preferred tactile-supported learning (\autoref{fig:ratings}(b)): 11 participants preferred \texttt{Tactile+Text+LLM}; one participant indicated that their preference depended on chart type complexity, and said tactile charts are more useful when learning more complex chart types; no participant preferred just \texttt{Text+LLM}. 
Participants also rated \texttt{Tactile+Text+LLM} as more useful for BLV education ($M = 4.83$) than \texttt{Text+LLM} ($M = 4.25$) (\autoref{fig:ratings}(c)). Consistent with these preferences, participants rated tactile charts as highly helpful for learning chart types ($M = 4.83$) and for subsequent new-dataset exploration with the LLM assistant ($M = 4.42$) (\autoref{fig:ratings}(d)). 
Participants viewed the LLM assistant positively, but rated it as less helpful for chart-type learning (\autoref{fig:ratings}(e), $M = 4.00$) than tactile charts (\autoref{fig:ratings}(d), $M = 4.83$). Participants rated the LLM assistant more helpful in exploring new datasets (\autoref{fig:ratings}(e), $M = 4.75$) than in chart type learning (\autoref{fig:ratings}(e), $M = 4.00$).

\section{Thematic analysis}
\label{sec:thematic-analysis}
We conducted a thematic analysis of the corrected transcripts~\cite{Lazar:2014:Research, Braun:2006:Using, Braun:2023:Toward}. The first author iteratively coded the data and developed themes through repeated engagement with the transcripts. Emerging codes and themes were discussed and refined with the research team. Below, we report the resulting themes with representative participant quotes. For readability, we removed filler words such as ``like'' and ``you know'' from quotes without changing the meaning of the sentences.
\rev{We also incorporated the analysis of LLM queries collected during the chart-type learning session into our thematic analysis, and report our analysis of queries generated while participants explored new datasets in \autoref{sec:queries-during-data-exploration}.}

\subsection{Chart-Type Understanding as a Prerequisite for Meaningful Exploration}

\begin{insightbox}
\rev{BLV participants described chart type understanding as a foundation for LLM-supported exploration.}
\end{insightbox}
Without an initial mental model of how an unfamiliar chart type is structured, participants found it difficult to interpret new datasets, ask questions of the LLM assistant and benefit from its responses. 
This finding highlights the importance of teaching participants unfamiliar complex chart types.

\textbf{Unfamiliar chart types are difficult to approach directly.}
Participants described exploring new datasets in unfamiliar chart types as cognitively demanding\rev{, and this challenge appeared across participants with different vision histories.}
For example, P12\rev{, who gradually lost their vision,} described dataset exploration without prior tactile learning as \hquote{The picture in your head may be a little bit difficult if this was a new concept}{P12-258}, and \hquote{if you were just kind of fresh and raw and green with it, then it, it could be very overwhelming very quickly for someone that's not familiar with how [the] data set and things are laid out.}{P12-261} 
Similarly, P5\rev{, who has been blind since birth,} noted that some concepts can be especially difficult to grasp, saying that \hquote{blind people, especially people who haven't seen at all before, struggle with the concept of shading, and densely populated [areas], or what those terms mean in a chart}{P5-916}. These comments suggest that before BLV individuals can effectively interpret data shown in complex visualizations, they first need access to the chart's conceptual logic.

\textbf{Knowing what to ask is essential for benefiting from AI.}
Especially in the context of using LLMs, participants also emphasized that LLM assistance was more useful for chart exploration when they already had some sense of what kinds of questions the chart could support. Understanding the chart type was often the foundation for asking good questions about the chart. 
P1 stated, \hquote{A better understanding when you ask good questions}{P1-329}, and emphasized, \hquote{so basically, know what you're gonna get out of it before you sit down and use the AI. But if you don't know what questions [to ask], you're gonna be lost.}{P1-549} 
P14 also emphasized the importance of asking the correct questions, noting, \hquote{I had to really think about how I wanted to ask the question so that the AI had an understanding of what I was looking for.}{P14-1120}
Similarly, P4 noted that the LLM assistant would be less useful for someone who \hquote{doesn't have more acquaintance with the chart}{P4-878}.

\subsection{Tactile Learning Builds Mental Models of Chart Types That Support Subsequent LLM Use}

Our prior work~\cite{He:2026:Using} showed that tactile charts can help BLV individuals learn unfamiliar chart types. This study confirms and extends that finding: participants reported that prior tactile learning helped them explore a new dataset using the same chart type with an LLM assistant.

\begin{insightbox}
\rev{BLV participants described the tactile learning experience as helpful for asking more targeted questions and better interpreting the LLM-based assistant’s responses.}
\end{insightbox}

\textbf{Tactile learning gave participants a chart template they could carry into later exploration.}
Participants repeatedly described the tactile chart as providing a general sense of how the chart type works, which they could reuse when encountering a new dataset in the same format.
For example, P1 explained that the tactile chart \hquote{basically helps your memory develop more---what is this chart about, what are we doing with it? And having the tactile thing, it kind of drills it into your memory as well, when it comes to cognitive processes.}{P1-525} 
P12 made a similar point, explaining that \hquote{having a tactile [chart] ahead of time, feeling it, I had a lot more [..] understanding of what it was going to look like, as opposed to the heat map [the condition in which the participant didn't have a tactile chart]. Having that tactile image gave me kind of a leg up on the violin versus the heat map.}{P12-391}
\rev{We can also see this pattern from participants with lower tactile literacy. For example, P9, who rated her familiarity with tactile charts or graphics as 2 out of 5, said that after learning with the tactile chart, she could approach the new dataset with the expectation that \hquote{If it was laid out, it would be in the same way.}{P9}}

For some participants, this prior tactile experience was especially important for making sense of chart features that were otherwise difficult to picture. 
P12 described the violin plot as \hquote{a foreign concept}{P12}, and said of tactile learning, \hquote{it did help in that next example to know, okay, this is what a violin model looks like}{P12}. 
Similarly, P14 said that learning with the physical violin plot helped: \hquote{so I had the understanding of the distribution and how it would appear}{P14-1212}. 
These comments suggest that tactile learning gave participants a more stable structural expectation before they entered LLM-supported data exploration.

\textbf{The tactile chart template supported more targeted question asking during LLM-based data exploration.}
As discussed above, knowing what to ask is important for making effective use of the LLM. Here, we focus on participants’ feedback on how tactile learning helped them ask more useful questions when later exploring a new dataset with the assistant.
This point is important because conversational exploration unfolds sequentially: what participants can get from the data depends in part on what they choose to ask. 
P4 explained this relationship especially clearly. After describing how the physical model helped him understand the chart's structure and purpose, P4 said, \hquote{Once you look into the charts physically, then depending on the explanation given from the AI, you can develop more questions. Because you have the understanding of the texture, you can already come up with some questions.}{P4-847} He also described an example scenario to show how tactile learning supported question formulation by giving him an internal reference for what features to probe further: After learning from the gentoo penguin's violin plot that a bimodal shape could reflect two subgroups, P4 felt able to ask targeted questions when encountering a new violin plot about which subgroups the peaks of the violin represented.

Other participants made similar connections between prior tactile learning and later question formulation with specific reasons. 
P7 noted that the tactile violin plot helped him understand the central purpose of the chart, explaining that \hquote{It imprints in your mind [..] the concept [of a] distribution; that the main thing is the fat distribution, and the median.}{P7-1080} When exploring the datasets with violin plots, he asked about these, e.g., \hquote{give me a broad overview and trends of the spread in all continents}{P7-query}, \hquote{take only the lowest half of each continent, then tell me the median}{P7-query}.
P2 highlighted tactile learning as supporting anticipation during iterative exploration. He explained that \hquote{having it there as a reference, gave you a better idea of what to expect, or to anticipate}{P2-325}. This reference also helped P2 feel more comfortable asking questions during subsequent exploration, as he noted: \hquote{I think I felt more comfortable asking questions, in the second [heatmap] example than I did in the second violin plot example, so I would say absolutely it did [..] help.}{P2-334} 

\rev{P9's experience further illustrates that this benefit could also appear for a participant with lower tactile-chart familiarity.} P9 said that tactile learning \hquote{helped me to be able to understand what to ask, and what kind of things---statistics---could be shown.}{P9-542} P9 did ask more specific questions related to statistics in clustered heatmaps (learned with tactile), e.g., extreme values, averages, and ranges in the heatmap, while in the violin plot, she only asked broad questions (e.g., asked for overall patterns, context) or non-targeted questions for information that is beyond the dataset.

\textbf{Tactile learning helped participants make sense of the assistant's responses.}
Beyond question asking, participants described the earlier tactile chart as a reference for understanding what the assistant said about the new dataset. P9 summarized this effect directly: \hquote{I kind of knew what kind of things to be listening for.}{}  
P2 made a closely related point, explaining that \hquote{having it there as a reference, gave you a better idea of what to expect, or to anticipate with, with the next [dataset]}{}, and concluded that \hquote{I think that did assist with assimilating the information more coherently.}{} 
P14 likewise described the first tactile example as helping her interpret the second dataset because \hquote{the first one helped me to explore the second [dataset], in that I had the concepts to work with. I had the previous experience of the concepts that I could then integrate and match up.}{P14}

\subsection{Using LLMs for Chart Type Learning}

In this section, we discuss participants' opinions on the benefits and limitations of the LLM assistant during chart-type learning. 
\begin{insightbox}
\rev{BLV participants valued LLM-based assistants for supporting interactive explanation, but these assistants could not replace tactile charts for spatial understanding. }
\end{insightbox}

\subsubsection{Benefits of LLM Assistants}
\label{sec:benefits-of-llm}
\textbf{LLMs supported a progressive, learner-driven process of clarification and elaboration.}
Participants described the value of the LLM assistant in terms of its conversational nature. Rather than delivering a fixed explanation, it allowed participants to begin with a partial understanding and then expand it through follow-up questions. As P5 said, 
\hquote{I think you'll get a general idea, and then using the AI kind of, like, expands everything.}{P5-969} P2, as an educator, tried to teach his students to use AI in their daily toolkit, because he believes \hquote{it has the potential to be very explanatory and to get granular when it needs to.}{P2-377} 

Participants particularly valued the LLM because it could clarify ideas when the provided materials were insufficient for learners' needs. 
\rev{For example, P1, who learned the violin plot in the \texttt{Text+LLM} condition, asked, \hquote{Can you give me a better explanation of the chart?}{} The chatbot responded, ``You can think of it as a mix of a boxplot (summary) and a smoothed histogram (shape).'' This analogy was not included in our pre-authored instructions. P1 found it helpful and commented, \hquote{Okay, I like the fact that when explaining with the AI, I didn't know the shape before, but now I was able to figure, oh, a histogram, okay, I've seen this before, so it kind of helps me more of explain‚  visualize in my head, okay, I know what it looks like now.}{} 
Similarly, P7 valued the LLM's ability to provide personalized explanations, stating, \hquote{The AI is very useful because you could basically ask it anything, and it can rephrase and reword things that the instruction may get funny, and you can ask it for additional context}{P7-1096}. He then further explained, \hquote{You can add anything, you can come at it from any angle. I think AI is more responsive and more, you know, it's like a back and forth, so you can take your own angle from it, which is good.}{P7-1102}}

In this sense, the LLM could complement the pre-authored materials (tactile charts and instructions) by helping participants address their personal confusion, which can be highly diverse and difficult for the pre-authored materials to cover. P1's interaction with an LLM in the tactile clustered heatmap condition illustrates this point. 
After asking why color was used to represent actors and requesting a clearer explanation, P1 continued by suggesting that textures might work better and explained that this would help \hquote{spot high and low values quicker}{P1-query}. 
After these turns of conversation with the assistant, 
P1 positively reflected on the interaction, saying, \hquote{I actually like it like this. They explain more of the chart with the AI when it comes to doing things like this, and it helps you more, okay, cool, it describes it better that way when you actually were able to ask it more questions like this.}{P1-253}
\rev{This learner-driven clarification may be especially useful for explaining complex chart components, such as dendrograms in clustered heatmaps. During learning, four participants asked dendrogram-related questions, in both the tactile (P5, P13) and non-tactile (P7, P14) conditions. Related queries included \hquote{Can you explain how the lines on the right side of the chart work?}{P5} and \hquote{does the dendrogram show relationships that are connected?}{P14}}

\textbf{LLMs could guide learning by suggesting next steps.}
\rev{Participants also noted that the LLM often supported learning by prompting the next step in the interaction.
Some follow-ups were directly prompted by the assistant, for example when it offered next steps and the participant chose one (e.g., the assistant ended their response with \hquote{If you tell me what you want to focus on---median, spread, or comparing groups---I can explain that part with a quick step-by-step reading strategy,}) and the participant accepted the offer). Others were clarification questions, such as \hquote{Do you mean the average for each species, or the overall average across all penguins in the CSV?}{P1}
Participants also initiated their own follow-ups by asking deeper questions based on earlier answers, such as continuing to probe how to interpret the dendrogram (P13's example above). This LLM behavior was observed in both learning and dataset exploration phases, but we considered it as part of the learning process and discussed it here because it helped participants deepen their understanding of the chart.}

P2 described this guided interaction process: \hquote{I was basically letting it lead me. Generally, in my interaction, what I've discovered is that it helps to have at least a general idea of what you're looking for, but then you can, based on what it tells you, become more granular in your analysis. You know, if you just treat it as organized brainstorming, essentially.}{P2-284}
P12 also described the LLM's prompts as invitations to dig deeper: \hquote{There was a couple [of instances] where it prompted me: Hey, do you want to know more? And I was like, yeah, tell me more, yes. [... It's like] digging down into that rabbit hole of Wikipedia, where you can click on one link, and then it's gonna take you to a whole new thing.}{P12-519}
P14 further emphasized that the assistant's response could redirect attention toward issues they had not initially considered: \hquote{So, for example, I could ask a specific question, and then his response would guide me to maybe an area I hadn't thought of before, or‚ would provide a suggestion which would extend [..] my understanding, because it would go and describe [it] in detail.}{P14-774}
P14 noted that \hquote{having AI almost be like a person you're having a conversation, and rather than having that person look at you kind of confused and being visually impaired, not understanding that that's what that meant, the AI is giving me the information: Hey what you asked isn't clear, so I need clarification, and is this what you're thinking about? And he gives examples, and then I could turn around and say, no, that's not exactly what I mean. So, it's like an ongoing conversation to broaden the understanding and to enrich the information.}{P14-1128}

\subsubsection{Issues with LLM Assistants}

Despite the benefits participants identified, they also described several limitations of the LLM assistant for chart-type learning.

\textbf{LLM explanations make it difficult to understand spatial structure.}
Like textual explanations, the LLM assistant could not fully convey spatial form and shape. Several participants in the non-tactile condition therefore expressed a need for tactile support. 
For example, P13, who learned the violin plot in \texttt{Text+LLM}, said while exploring the new dataset: \hquote{I struggled to get an actual picture of how many curves, I suppose, the violin has. The alt text gave an average, and the AI wasn't really able to say, oh, there are 10 points at which the violin, you know, expands on Asia or Africa. So I think for that, I would need to physically see it.}{P13 430}

\textbf{Effective use of LLMs required some AI literacy and prompting skill.}
Participants also emphasized that the usefulness of the LLM assistant depended partly on knowing how to interact with it. 
P2 described this as a learning curve, explaining, \hquote{there are the natural limitations that occur when there is learning how to ask the questions in the right way. Sometimes you get the old equation: garbage in, garbage out. If I don't ask the question the right way, I'm not going to get an answer that is coherent. There's a small learning curve.}{P2-340}
At the same time, P2 is positive about LLMs, saying, \hquote{I think that that learning curve, every year, is going to become exponentially smaller as people become more accustomed to working with AI, looking over their shoulder.}{P2-340} P2 noted that the training in this study was sufficient: \hquote{I think by the time you get to the second data set, [..] most people will have become sufficiently comfortable that they're better able to come out of the gate asking the right questions, or at least questions that are better designed to elicit a more coherent response from the chatbot.}{P2-352} P4 shared this opinion, 
\hquote{[..] The helpfulness of the AI really differs. It depends on the skill of the individual on how to use it, and on how they are familiar [..]. The more familiar you are with the AI and with your tools, you can ask more questions, you can get more help. If you are less familiar, it might not even be helpful.}{P4-861}
P4 also believed that users’ educational backgrounds could affect their ability to formulate effective prompts and, consequently, obtain useful responses.

\textbf{Participants worried about trust and information overload.}
Participants identified two broader challenges in using LLMs for chart learning: the risk of inaccurate responses and the difficulty of processing too much numeric information. These concerns appeared mainly during dataset exploration, but also surfaced in chart-type learning. For example, P7 emphasized that the assistant is most useful when its answers are grounded in the dataset and can be trusted, saying, \hquote{if it's trained with a specific data set like this. I think it is especially useful, because it doesn't give you random answers, it is already trained, and as long as we can verify that it's not hallucinating, I think it is extremely useful when it gets to the point.}{P7-1117}
Several participants also noted that responses with too many numbers were hard to follow. For example, P5 said she would prefer the LLM to \hquote{figure out how to consolidate some of the information so it's not saying the numbers so rapidly, and, one after the other.}{P5-891} 

\rev{\textbf{LLM assistants need to provide more adaptive scaffolding that helps learners recognize and overcome gaps in understanding.}
Participants noted that LLM assistants need to be more sensitive to learners’ prior knowledge. P4 suggested that, ideally, AI systems should detect knowledge gaps and adjust the depth and pace of explanations. P14 suggested, \hquote{The chatbot shouldn't just have assumptions that‚ when you go into this, you already know A, and so, it can jump to B. It might need to [..] check for knowledge of A before jumping ahead to B.[..] If I don't have that [..] the chat bot [should] be able to modify its response to reflect that.}{P14-1285} P4 later clarified that he meant both experience with the chart types and with using AI assistants.}

\rev{Beyond modeling learners' knowledge, we also observed the need for LLMs to help learners recognize what they do not yet understand. 
Because learning with an LLM is largely learner-driven, novices may struggle to identify their own knowledge gaps and propose appropriate questions. 
Therefore, LLM assistants should provide stronger scaffolding. 
In the chart learning phase, we observed that four participants (P3, P4, P11, P12) did not ask any questions while learning either chart, and five participants (P1, P5, P10, P13, P14) asked questions for only one of the two charts. Most of these participants reported that the provided instructions were sufficiently clear. However, their later responses to the comprehension questions were not always correct, which suggests that their understanding of the chart types could still be improved.
P10's experience illustrates this challenge clearly.
Although she found the violin plot difficult to understand, she explained that she did not ask the assistant questions \hquote{because I don't know what to ask it}{P10-409}. She further noted that, even with a human instructor, it would still have been difficult to \hquote{figured out how to even know what to ask}{P10}.
For such learners, the LLM assistant may need to take a more proactive role by offering possible questions related to this chart type, or asking low-stakes diagnostic questions to check learners' comprehension.}

\rev{LLM assistants also need to better detect the underlying reason for a learner’s confusion from their questions, and provide explanations that directly target it.
P13’s experience with dendrograms was particularly illustrative. He first asked, \hquote{Do all heatmaps have dendrograms?}{P13}, then followed with more specific questions about interpreting the structure: \hquote{It's definitely clustered. Do the distances between dendrograms also provide information, or are they separated linearly?}{P13}, \hquote{I see several areas where the dendrograms don't join with a row. Is that normal? There are other places where two lines are obviously connected,}{P13} and \hquote{Let's look at Dwayne. His dendrogram extends far out and appears to eventually make contact with almost everyone else—save for Julia Roberts. Why?}{P13} However, even after reading the AI’s responses, P13 still did not understand the hierarchical structure. He explained, \hquote{I'm not exactly clear on why it doesn't look like some of these connected, because my understanding would be you should have a connection from every row to every other row.}{P13} The interviewers then explained that the dendrogram represents stepwise grouping: the two most similar actors first form small groups (Julia Roberts and Tom Hanks, Jennifer Lawrence and Leonardo DiCaprio), which then merge into larger groups. P13 immediately reformulated this correctly: \hquote{And then Dwayne connects to the group of all four, because he's so far out there,}{P13} and concluded, \hquote{That makes sense. Now I think I see what I was missing.}{P13} In this case, the LLM did not fully resolve P13’s confusion, likely because it provided too many technical details about hierarchical clustering and multiple possibilities without addressing his core misunderstanding: that dendrograms show groups merging step by step, rather than each row directly connecting to others.}

\subsection{Multimodal Chart-Type Learning}

In our prior work \cite{He:2026:Using}, participants preferred a combination of tactile charts with structured exploration instructions for chart-type learning over text-only learning. In this study, we asked whether an LLM assistant, with its interactivity and flexibility, can compensate for the shortcomings of static text, leading participants to prefer it over tactile charts. We found that participants did not consider the assistant as a replacement for tactile charts. Instead, they described the best learning experience as a combination of tactile charts, exploration instructions, and the LLM assistant.
\begin{insightbox}
\rev{BLV participants preferred a multimodal learning experience for chart type learning, because tactile charts, text, and LLMs provided complementary forms of support.}
\end{insightbox}

\textbf{Participants preferred multimodal learning because each modality contributed a different kind of understanding.}
P2 highlighted the importance of reinforcement across modalities, saying that \hquote{using as many different senses as possible to aid in comprehension is always, I think, a bonus.}{P2-319}
P14 also noted, \hquote{Just the AI is just an auditory/visual modality. I don't really get to incorporate multiple senses.}{P14-1158} She described the combined setting as:  
\hquote{Three different ways of looking at the information, and each one being its own means of providing a different or adding dimension to it.}{P14-1172} 
P7 emphasized the value of multimodal learning because of the needs of different people, saying, \hquote{I think people learn in different ways. I think for a lot of blind people, tactile learning is probably one of the best, of course, and AI is interactive, it's good!}{P7-1115}
P4 further shared his strategies for using these modalities together: \hquote{Usually, for me, I'd love to have the physical chart, and then go to the AI whenever I need more explanation, and I believe that more work can be done on the physical chart as well.}{P4-820}
P7 also envisioned a more interactive design to combine these modalities: \hquote{AI is connected to the refreshable graphics board, and that graphics board can change based on AI, and AI can draw the line for you from place to place if you ask. If you ask about the trend of, [or] about the peak of something. It can have a moving dot there to show your finger where it is, and maybe even show a number there to describe to you numbers.}{P7-1148}

\textbf{When tactile support is unavailable, LLMs can partially compensate for its absence and provide greater support than text alone.}
Although participants preferred multimodal learning, they also recognized the practical value of LLM assistance.
P7 framed this explicitly as a real-world constraint, noting that \hquote{In real life, it's going to be a while until we can do the [3D] print on a whim. So, I think a lot of blind people, including myself, are going to be quite happy with text description and an AI chatbot, if we can make it this accurate}{P7-1168}. 
He further stated that the combination of text and LLM \hquote{is better than text only.}{P7-1129} 
P1’s experience in the \texttt{Text+LLM} condition shows how the assistant could partly compensate for missing tactile support: as we discussed in \autoref{sec:benefits-of-llm}, the LLM’s explanation that the violin plot is similar to a histogram helped him understand the chart.

\section{Discussion}
Our findings suggest that BLV participants preferred a multimodal learning experience: across the interviews and subjective ratings, participants valued \texttt{Tactile+Text+LLM}, because each modality contributed differently to the learning process. Tactile charts were especially important for conveying spatial information, which is difficult to communicate through other modalities. This spatial understanding appeared to provide an important foundation for subsequent data exploration. The textual introduction helps participants understand the tactile model and sets up the dataset. LLMs, in turn, were valued for their flexibility. Compared with static descriptions such as alt text or instructions, LLMs can respond to a learner's immediate confusion, elaborate on partially understood concepts, support exploration through dialogue, rely on ``world knowledge'' for broader questions, and answer queries based on patterns in the data that may not be encoded in a visualization. This flexibility seems especially useful once learners have formed an initial mental model and want to refine or extend their understanding. 

The value of LLM support is also reflected in comparison with our previous work~\cite{He:2026:Using}. Although the two studies involved different participants, both targeted the same population and used the same chart types and new datasets. Participants in the present study reported higher perceived understanding of the complex datasets, with mean ratings ranging from 3.83 to 4.33 under the \texttt{Tactile+Text+LLM} and \texttt{Text+LLM} conditions, compared with 3.50 to 3.83 under the \texttt{Tactile+Text} and \texttt{Text-Only} conditions in the prior work. Similarly, perceived usefulness in education was higher for \texttt{Text+LLM} ($M = 4.25$) than for \texttt{Text-Only} ($M = 4.00$). 

In addition, LLMs have the practical advantage of being immediately available as a digital support tool and, obviously, have enormous flexibility across a wide range of contexts. However, they still cannot convey what a chart feels or looks like. Therefore, we suggest using multimodal systems to support BLV chart learning, especially when formally educating BLV individuals about data and data visualization, such as in college classes. These modalities should be closely connected. In particular, LLM support for chart learning should be grounded in both the tactile chart and the underlying data, so learners can smoothly move between physical exploration and conversational explanation.

Although participants' answers to factual observation questions, their subjective ratings, and their interview comments all pointed in a more positive direction, our quantitative accuracy measures did not show an improvement for \texttt{Tactile+Text+LLM} on chart understanding questions. 
\rev{This mismatch between perceived value and measured performance could have several potential reasons. First, the effect of tactile chart type learning on subsequent data exploration may be difficult to detect with our small sample size. 
Second, the chart type learning phase was relatively short, lasting 18 $\pm$ 6 minutes. It is possible that short-term chart type learning does not strongly transfer to measurable performance improvements during data exploration. Future work could examine this possibility through longer-term studies, for example in collaboration with BLV educational organizations.
Third, the questions we asked may not have been the most suitable measures for capturing learning outcomes or participants’ mental models of the chart type.}
This limitation points to an important methodological opportunity for future work: to develop evaluation methods that better capture how BLV individuals benefit from visualizations and non-visual representations in supporting data analysis.

Our findings also suggest that the LLM assistants in this study may not always be sensitive enough to learners' prior knowledge. For example, when explaining complex components such as dendrograms, the LLM became lost in technical details, whereas a human interviewer could intervene more effectively because they were better able to recognize the source of the difficulty for participants in understanding the chart. In this study, we intentionally limited customization mostly to the system-prompt level in order to approximate how LLMs might be used in practice. We did not provide explicit instruction-customization options, as prior work showed that participants with similar backgrounds rarely customized the assistant themselves \cite{Seo:2024:MAIDRAI}. However, our findings suggest that future LLM systems for BLV chart learning could benefit from more adaptive behavior, such as detecting knowledge gaps during conversation, accurately identifying where participants are getting stuck, and adjusting explanations accordingly. 

\section{Conclusion and Future Work}

\rev{Our thematic analysis suggests that tactile charts and LLMs complement each other in supporting BLV individuals’ learning of unfamiliar complex chart types: tactile charts provide essential spatial understanding, while LLMs support flexible clarification and follow-up exploration. This combination points to a promising multimodal approach for improving chart learning and thereby enabling better participation in data analysis by BLV individuals.}

Our study focused on short-term learning in a controlled setting rather than long-term learning in real educational contexts. However, participants recognized the potential value of these learning modalities in real educational settings. As one participant noted about \texttt{Tactile+Text+LLM}, \hquote{I wish I had this when I was in school, that would have really helped out.}{P1} 
This response suggests promising directions for future research on applying these learning formats in real educational settings, including schools for BLV students and other real-world learning environments.

Participants also expressed interest in making tactile access more adaptive. This points to promising opportunities for future work on refreshable tactile displays combined with LLM assistants, which could provide a more responsive and interactive learning experience than static tactile materials, although cost and access remain important practical considerations.

\acknowledgments{%
The authors thank Aaron Delahunta and Sofia Djunic for correcting interview transcripts. 
}

\bibliographystyle{abbrv-doi-hyperref-narrow}
\bibliography{abbreviations,processed-output}

\appendix 
\crefalias{section}{appendix} 

\clearpage

\begin{strip} 
\noindent\begin{minipage}{\textwidth}
\makeatletter
\centering%
\sffamily\bfseries\fontsize{15}{16.5}\selectfont
\papertitle \\[.5em]
\large Appendix\\[.75em]
\makeatother
\normalfont\rmfamily\normalsize\noindent\raggedright 
In this appendix, we provide additional information and figures that we could not include in the main paper due to space limitations.
\end{minipage}
\end{strip}

\section{Interview Questions}
\label{sec:interview-questions}

\subsection{Background Questions}
\label{sec:background-questions}
\subsubsection*{Demographics}

\begin{itemize}
    \item If you would like, please let us know your preferred pronouns.
    \item Year of birth
    \item Gender
    \item What is the highest level of education you have completed?
    \item What is your current status? Are you working or studying, and is it full-time or part-time? If so, what is your job or major?
\end{itemize}

\subsubsection*{Vision loss level}

\begin{itemize}
    \item How would you describe your vision-loss level?  (Blind since birth / Lost vision suddenly / Lost vision gradually)
    \item How would you describe your vision level (e.g., completely blind, light perception, central vision loss, etc.)?
    \item (If the participant is not totally blind) If you know, what is your corrected visual acuity in either Snellen (e.g., 20/200) or LogMAR (e.g., 1.3)?
    \item If you are comfortable sharing, please indicate your visual pathology diagnosis. This information is optional and not required.
\end{itemize}

\subsubsection*{Screen Reader Experience}
\begin{itemize}
    \item Which screen reader do you use with your computer or other devices (e.g., NVDA, JAWS, VoiceOver, etc.)?
    \item How long have you been using a screen reader?
    \item When using a screen reader, what is your preferred rate of speech?
    \item Do you use other accessibility devices or software in combination with a screen reader, such as screen magnification or a Braille display? If yes, please describe all accessibility devices or software you use in combination with a screen reader.
\end{itemize}

\subsubsection*{Braille experience}

\begin{itemize}
    \item How long have you been reading Braille?
    \item How would you rate your skill in reading Braille (1 - I can't read Braille / 5 - I'm very proficient at reading Braille)
    \item In what contexts did you read Braille (e.g., work, personal use)?
\end{itemize}

\subsubsection*{Tactile Chart Experience}
\begin{itemize}
    \item How would you describe your familiarity with tactile charts or graphics? (1 - Not at all familiar / 5 - Extremely familiar)
    \item Can you describe in which contexts you have interacted with tactile charts or graphics?
\end{itemize}

\subsubsection*{LLM experience}
\begin{itemize}
   \item Have you ever used an AI chatbot assistant or a large language model (LLM), like ChatGPT, Google Gemini, Claude, or something similar?
   \item Which tool or tools do you use?
   \item How often do you use these tools?
   \item What do you usually use them for?
   \item Have you used an AI assistant to interpret data—like a dataset, table, chart, or visualization?
\end{itemize}

\subsubsection*{Data Visualization Experience}
\begin{itemize}
    \item How many hours do you use a computer or smartphone each day?
    \item Would you consider your career to be data-intensive or numbers-driven (e.g., regularly work with large datasets, perform statistical analyses, or make decisions based on quantitative information)?
    \item How often do you interact with data visualizations, such as those for work, from news articles, in video games, etc.?
    \item In which contexts do you encounter data and data visualizations (e.g., work, news, leisure)?
    \item How do you typically view datasets? (e.g., download to excel, you don't, etc.)
\end{itemize}

\subsubsection*{Familiarity With the Two Chart Types}
\begin{itemize}
    \item How familiar are you with \textbf{clustered heatmaps} - a visualization of tabular data that encodes values by color? (1 - Not at all familiar / 5 - Extremely familiar)
    \item Have you seen or touched \textbf{clustered heatmaps} through a tactile display or an embossed paper before? (Yes / No / I don't know what a clustered heatmap is)
    \item How familiar are you with \textbf{violin plots} - a visualization of a distribution of values? (1 - Not at all familiar / 5 - Extremely familiar)
    \item Have you seen or touched \textbf{violin plots} through a tactile display or an embossed paper before? (Yes / No / I don't know what a violin plot is)
    \item After you received our package, have you explored the model, or heard any of the instructions?
\end{itemize}

\begin{itemize}
    \item Researcher B, do you have any additional questions?
    \item To ensure that we are both on the same link during today's discussion, do you mind also sharing your screen?
\end{itemize}

\subsection{Questions After Chart Type Instructions}
\label{sec:questions-chart-type-instructions}
\begin{itemize}
    \item Can you describe what a violin plot / clustered heatmap is, in your own words?
    \item What types of data are best suited for visualization using a violin plot / clustered heatmap?
    \item Can you provide an example scenario where you would use a violin plot / clustered heatmap?
\end{itemize}

[Only For Violin Plot]
\begin{itemize}
    \item If a violin’s shape has three peaks, what does it mean for the underlying data?
    \item If two distributions have the same median and overall range but different spreads around the center (one is tightly clustered, one is more dispersed), how would their violins differ in shape?
\end{itemize}

[Only For Clustered Heatmap]
\begin{itemize}
    \item In a clustered heatmap, if two rows (or columns) are very similar, how would you expect them to appear in the heatmap and in the dendrogram?
    \item If one row (or column) looks very different from all the others, how would that show up in the heatmap and in the dendrogram?
\end{itemize}

\subsection{Questions After Complex Dataset Exploration}
\label{sec:questions-complex-alt-text}
\begin{itemize}
    \item What is the dataset about?
    \item Imagine you have a friend who has never encountered this chart before. How would you describe this chart to them?
    \item What are the numerical characteristics being represented by this chart?
    \item What are the trends represented by this chart?
    \item What did you find interesting or surprising about the dataset?
\end{itemize}

\subsection{Comparative Questions After Both Charts Were Explored}
\label{sec:comparative-questions}

\begin{itemize}
    \item You have learned about visualization types in two different ways: one by exploring a tactile model with instructions and the LLM, and another by only hearing a textual explanation and the LLM. Which training format do you prefer? 
    \item How helpful is the tactile model used in the training for understanding the chart type?
    \item Do you think the tactile model helped you explore and interpret the new dataset  (the second dataset for each chart type) with the AI assistant? 
\item How helpful is the AI assistant for interpreting new datasets (the second dataset for each chart type)?
    \item How helpful is the AI assistant used in the training for understanding the chart type?
\item How useful is the training format “tactile model with exploration instructions and AI assistant” for BLV education in your opinion?
\item How useful is the training format “textual explanation and AI assistant” for BLV education in your opinion?
\item Do you have any other feedback or comments that we didn’t touch on today?
 \end{itemize}

\clearpage

\section{Simple Alt Text}
\label{sec:simple-alt-text}
\subsection{Clustered Heatmap}

This is a \textbf{clustered heatmap} showing how frequently actors appear in different movie genres.

\begin{itemize}
    \item \textbf{Rows:} Actors.
    \item \textbf{Columns:} Movie genres.
    \item \textbf{Cells (color intensity):} The number of appearances for each actor–genre pair. \textbf{Darker} cells indicate \textbf{more} appearances, and \textbf{lighter} cells indicate \textbf{fewer} (or none).
    \item \textbf{Clustering:} Actors an\textbf{d genres are \textbf{clustered by similarity}. A hierarchical tree (dendrogram)} next to the rows shows which actors cluster together (shorter branch distances indicate more similar genre-appearance patterns). Another dendrogram above the columns shows which genres cluster together.
\end{itemize}

\textbf{Overall pattern:}
\begin{itemize}
    \item Julia Roberts and Tom Hanks show similar patterns, focusing primarily on Drama and Comedy
    \item Jennifer Lawrence and Leonardo DiCaprio share a strong emphasis on Drama
    \item Dwayne Johnson is an outlier in this dataset, with a strong specialization in Action and no Romance films
\end{itemize}

\subsection{Violin Plot}

This is a violin plot showing the distribution of body mass (in pounds) across three species of penguins: Adelie, Chinstrap, and Gentoo.

\begin{itemize}
    \item \textbf{Y-axis:} Body mass (pounds).
    \item \textbf{X-axis:} Species, from left to right: \textbf{Adelie, Chinstrap, Gentoo}.
    \item \textbf{Violins:} Each violin shows the distribution of body mass for that species.
\end{itemize}
\textbf{Overall pattern: }Adelie and Chinstrap have unimodal distributions with similar median values. Gentoo shows a bimodal distribution and a higher median than the other two species.

\section{Complex Alt Text}
\label{sec:complex-alt-text}

\subsection{Clustered Heatmap}

This is a \textbf{clustered heatmap} showing how citizens in \textbf{15 European countries} rate the relative importance of \textbf{9 sociocultural, institutional, and religious topics} (for example, \textbf{confidence in government}, \textbf{belief in God}, and \textbf{attitudes toward homosexuality}) based on a survey.

\begin{itemize}
    \item \textbf{Rows:} Topics.
    \item \textbf{Columns:} Countries.
    \item \textbf{Cells (color intensity):} The agreement level for each country--topic pair. Darker cells indicate higher agreement, and lighter cells indicate lower agreement.
    \item \textbf{Clustering:} Both countries and topics are \textbf{clustered by similarity}. A \textbf{hierarchical tree (dendrogram)} next to the rows shows which topics cluster together (shorter branch distances indicate more similar response patterns). Another dendrogram above the columns shows which countries cluster together.
\end{itemize}

\textbf{Overall pattern:} The Nordic countries are the most confident in state institutions and most liberal. Religious values are strongest in Russia, Italy, Poland, Germany, Greece and Portugal. Poland, Italy, and Russia are most conservative on social issues.

\subsection{Violin Plot}

This is a violin plot showing the distribution of the Human Development Index (HDI) for countries in 2021, grouped by continent: Africa, Americas, Asia, Europe, and Oceania.

\begin{itemize}
    \item \textbf{Y-axis:} HDI, ranging from 0 (low) to 1 (high).
    \item \textbf{X-axis:} Continents.
    \item \textbf{Violins:} Each violin shows how countries' HDI values are distributed within that continent (wider sections indicate more countries at that HDI level). A central marker indicates the median.
\end{itemize}

\textbf{Overall pattern:} Europe exhibits the highest and most consistent HDI values, while Asia has the widest variability. Africa and Oceania have non-normal distributions, indicating groups of countries with different levels of development. HDI is measured from 0 (low) to 1 (high).

\clearpage

\section{\rev{Queries During Data Exploration}}
\label{sec:queries-during-data-exploration}

We collected 276 queries in total. Of these, 13 were practice queries used by participants to familiarize themselves with the chatbot at the beginning of the study (e.g., ``Hello'') and were not related to the charts. The remaining 263 queries were made during chart learning or data exploration and were used in the analysis reported in this section. In some cases, queries were not logged due to technical issues, such as browser refreshes. We recovered these queries from the Zoom video recording and included them in the dataset.

\rev{In this section, we analyze the queries collected from the complex alt text phase. We incorporated the analysis of queries collected during the chart-type instruction phase into our thematic analysis, reported in \autoref{sec:thematic-analysis}.}

We collected 140 queries from the complex alt text phase, where participants explored the new datasets across the two chart types. Of these, 138 were substantive, excluding responses such as ``thank you.'' Because the tactile model did not encode the new data, the types and frequencies of exploration queries were similar across conditions, so we do not distinguish them here.
Kim et al.~\cite{Kim:2023:Exploring} characterized BLV participants’ chart-QA queries in a Wizard-of-Oz study using prepared answer sheets. Here, we focus on patterns not covered in their analysis, such as interactions with a real LLM, while also comparing our results with theirs on points most relevant to our goals, especially chart-type learning.

\parhead{Chart-/Dataset-Answerable Questions}
Of the 138 substantive queries, 99 could be answered using the chart and the dataset we provided to the LLM. 
Thirty queries needed to be answered based on the dataset, using information that was not directly encoded in the chart. 
This was especially common in the violin plot, where the dataset contains country-level information, whereas the chart shows only continent-level distributions. Some participants asked country-level questions, such as \hquote{List the highest and lowest countries HDI [human development index]}{P10}, or \hquote{compare contrast northern and southern Europe for me}{P7}. One participant also directly asked the LLM to provide the dataset---\hquote{Please output the table in CSV format}{P13}---because he wanted to analyze it in Excel. For some queries that were answerable from the chart, the LLM relied on the dataset to produce a more precise response. For example, for a query such as \hquote{take the lowest half of Europe and the top half of Africa, contrast and compare, give me any interesting insights}{P7}, the assistant tended to support its answer with exact values such as ranges or medians, drawn from the dataset. This highlights a strength of LLMs that have access to data: unlike a predefined chart, the LLM can answer more detailed questions users may have. 

\parhead{Questions Leveraging LLM's World Knowledge}
For queries that were not answerable from the chart or the underlying dataset, several were answerable with the LLM's world knowledge. These included requests for definitions (e.g., \hquote{How is HDI [human development index] determined?}{P13}), underlying reasons (e.g., \hquote{why do Northern European countries have such a high HDI}{P12}), and perspective-based interpretations (e.g., \hquote{tell me more about this data set from a christian perspective?}{P12}), again showing the benefits of using an LLM that can incorporate world knowledge beyond what even an interactive visualization or the full dataset could reveal.  

\parhead{Out-of-Scope Questions}
We observed 11 queries about details beyond the dataset, which Kim et al.~\cite{Kim:2023:Exploring} characterize as out of scope. Most of these came from P3 and P12. For example, P3 asked for information not contained in the dataset, such as \hquote{how does population data work in this chart}{P3}, or \hquote{how is it determined which country is conservative}{P3}. P12 asked three questions related to Kansas, e.g., \hquote{How does the HDI of Kansas compare to the USA HDI?}{P12}, but he later explained that he simply wanted to test whether the LLM was limited to the provided dataset or could use outside knowledge.
In these cases, the LLM typically acknowledged the limitation in its response (e.g., ``I can’t tell from the dataset I have: it only includes country, Human Development Index (value), and region---there’s no population column here.'') and sometimes suggested uploading additional data.

\parhead{Unclear Questions}
We observed 10 unclear queries. Some involved genuinely ambiguous references, similar to those reported by Kim et al.~\cite{Kim:2023:Exploring}. For example, in \hquote{when it comes to the chart, give me the data from high to low}{P1}, it is unclear what ``data'' refers to, so the LLM asked for clarification and suggested possible interpretations. Other unclear queries reflected intuitive expressions that made sense to participants but required further specification for the LLM. For example, in \hquote{which continents have a similar HDI}{P13}, the LLM requested a more explicit definition of ``similar'', such as similarity in mean or range. 

\parhead{Follow-Up Questions}
During dataset exploration, some participants relied more heavily on the LLM's guidance and followed the LLM through multi-turn exchanges. For example, while exploring the violin plot, P5 replied \hquote{Yes}{P5} three times to continue along the LLM’s suggested directions.


\clearpage

\section{Tactile Charts}
\label{sec:tactile-charts}
\begin{figure}[!h]
    \centering
        \includegraphics[width=1\columnwidth, alt={An overhead photograph of a 3D-printed tactile clustered heatmap. A five-row by four-column matrix of raised cells is accompanied by dendrograms above and to the right. Braille labels and a Braille key surround the chart. A clipped upper-right corner indicates its orientation.}]{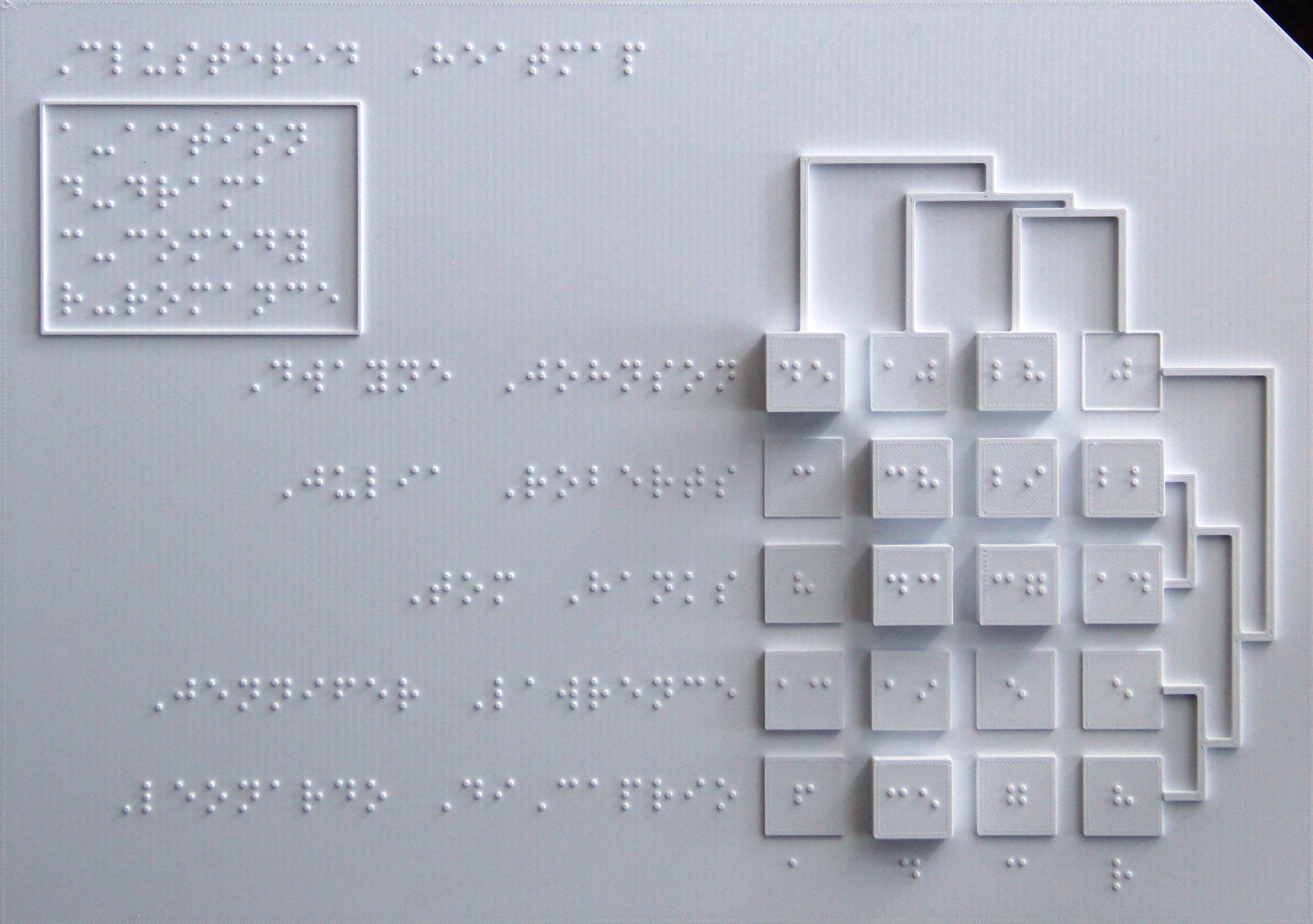}
    \caption{The 3D-printed tactile chart for the clustered heatmap, front view.}
    \label{fig:heatmap-final-front}
\end{figure}

\begin{figure}[!h]
    \centering
        \includegraphics[width=1\columnwidth, alt={An overhead photograph of a 3D-printed tactile violin plot. Three raised violin shapes appear on a grid with tactile markers for median values. Braille labels identify the axes and categories. A clipped upper-right corner indicates its orientation.}]{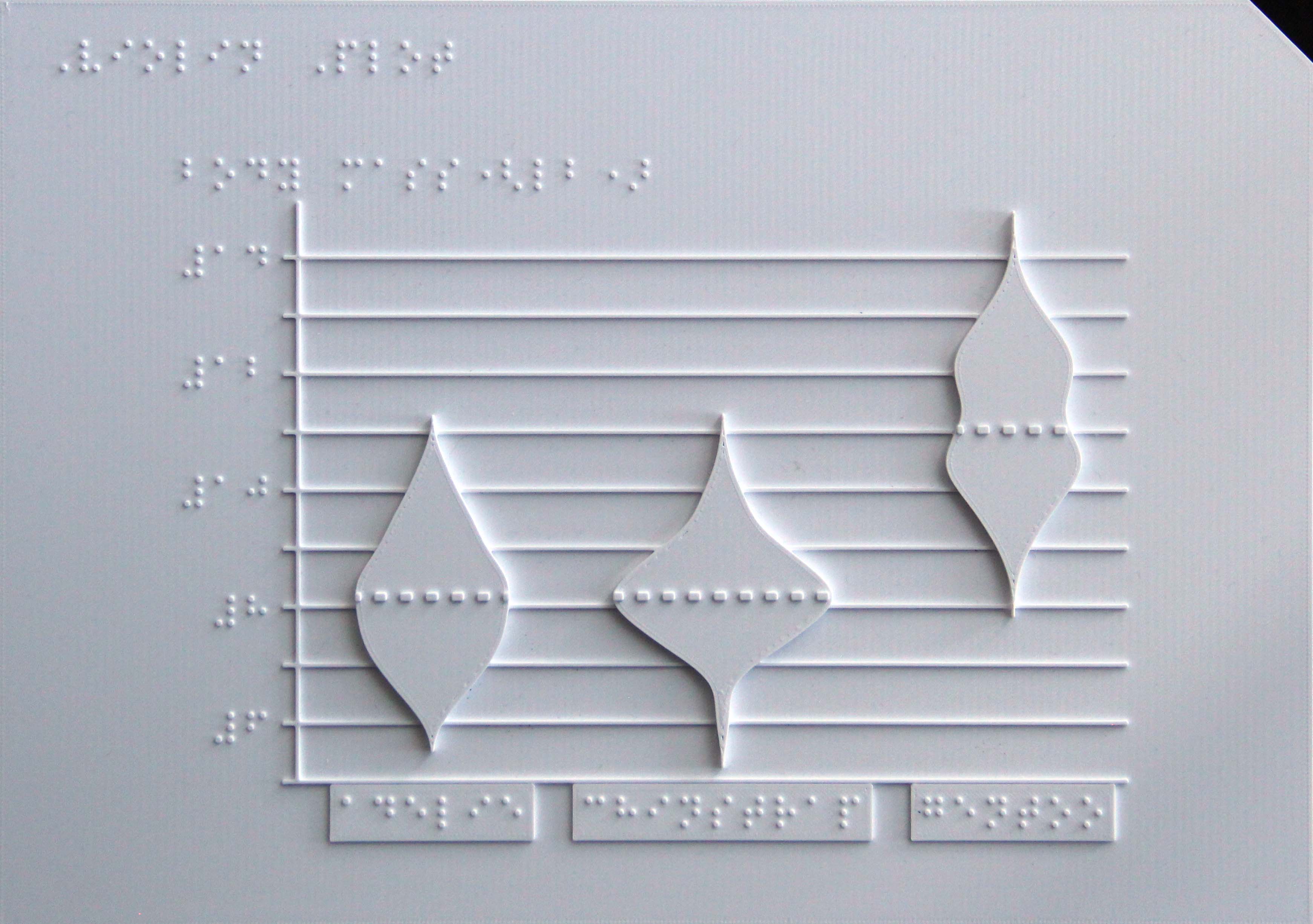}
    \caption{The 3D-printed tactile chart for the violin plot, front view.}
    \label{fig:violin-final-front}
\end{figure}    

\newpage

\section{Complex Charts}
\label{sec:complex-charts}
\begin{figure}[!h]
    \centering
        \includegraphics[width=1\columnwidth, alt={A grayscale clustered heatmap of responses from 15 European countries to nine sociocultural, institutional, and religious topics. Countries are columns and topics are rows. Darker cells indicate higher agreement. Dendrograms above and to the left group countries and topics with similar response patterns.}]{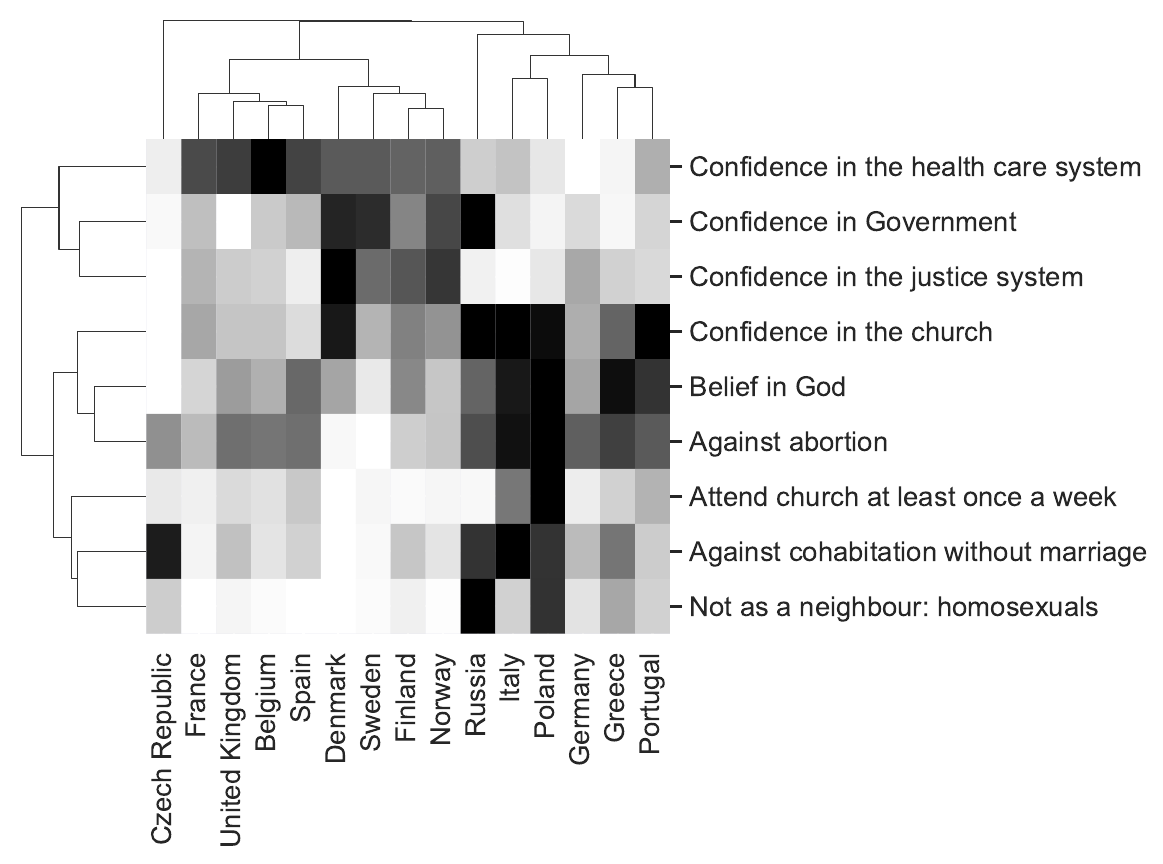}
    \caption{The clustered heatmap with the complex dataset.}
    \label{fig:heatmap_complex}
\end{figure}    

\begin{figure}[!h]
    \centering
        \includegraphics[width=1\columnwidth, alt={A violin plot showing the distributions of Human Development Index values across five continents. Dashed lines within each violin indicate the quartiles and median.}]{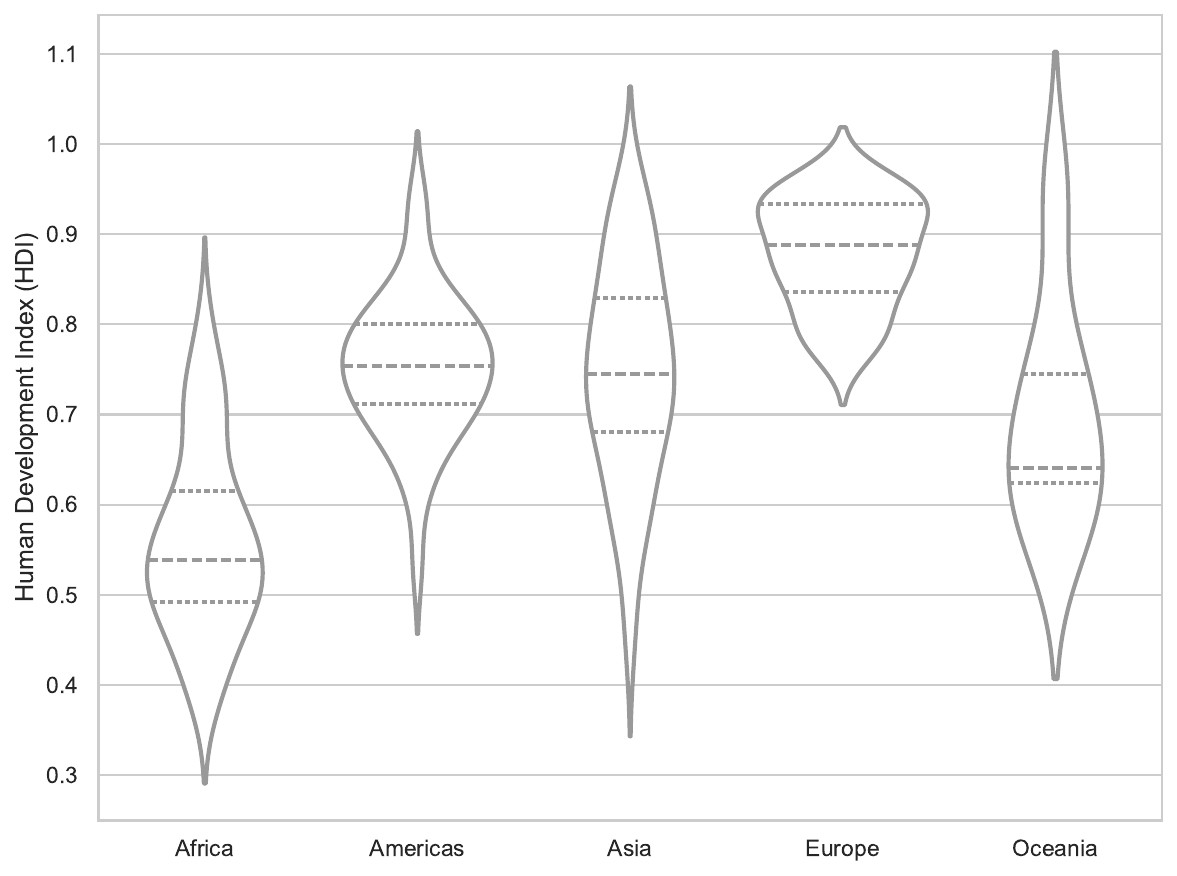}
    \caption{The violin plot with the complex dataset.}
    \label{fig:violin_complex}
\end{figure}

\end{document}